\documentstyle[12pt,epsfig]{article}
\newcommand{\myfigure}[3]{ 
\begin{figure}[tbhp]
\begin{center}
\mbox{
      \epsfig{file=#1,width=#2\textwidth}
      }
 \caption{#3}
 \end{center}
\end{figure} 
}

\begin{document}

\newskip\humongous \humongous=0pt plus 1000pt minus 1000pt
\def\caja{\mathsurround=0pt}
\def\eqalign#1{\,\vcenter{\openup1\jot \caja
 \ialign{\strut \hfil$\displaystyle{##}$&$
 \displaystyle{{}##}$\hfil\crcr#1\crcr}}\,}
\newif\ifdtup
\def\panorama{\global\dtuptrue \openup1\jot \caja
 \everycr{\noalign{\ifdtup \global\dtupfalse
 \vskip-\lineskiplimit \vskip\normallineskiplimit
 \else \penalty\interdisplaylinepenalty \fi}}}
\def\eqalignno#1{\panorama \tabskip=\humongous
 \halign to\displaywidth{\hfil$\displaystyle{##}$
 \tabskip=0pt&$\displaystyle{{}##}$\hfil
 \tabskip=\humongous&\llap{$##$}\tabskip=0pt
 \crcr#1\crcr}}
\jot = 1.5ex
\def\baselinestretch{1.2}
\parskip 5pt plus 1pt
\catcode`\@=11
\@addtoreset{equation}{section}
\def\theequation{\thesection.\arabic{equation}}
\def\@normalsize{\@setsize\normalsize{15pt}\xiipt\@xiipt
\abovedisplayskip 14pt plus3pt minus3pt%
\belowdisplayskip \abovedisplayskip
\abovedisplayshortskip \z@ plus3pt%
\belowdisplayshortskip 7pt plus3.5pt minus0pt}
\def\small{\@setsize\small{13.6pt}\xipt\@xipt
\abovedisplayskip 13pt plus3pt minus3pt%
\belowdisplayskip \abovedisplayskip
\abovedisplayshortskip \z@ plus3pt%
\belowdisplayshortskip 7pt plus3.5pt minus0pt
\def\@listi{\parsep 4.5pt plus 2pt minus 1pt
     \itemsep \parsep
     \topsep 9pt plus 3pt minus 3pt}}
\relax
\catcode`@=12
\evensidemargin 0.0in
\oddsidemargin 0.0in
\textwidth 6.0in
\textheight 8.5in
\hoffset .7 cm
\headsep .75in
\topmargin=-1in
\catcode`\@=11
\def\section{\@startsection{section}{1}{\z@}{3.5ex plus 1ex minus
   .2ex}{2.3ex plus .2ex}{\large\bf}}

\def\thesection{\arabic{section}}
\def\thesubsection{\arabic{section}.\arabic{subsection}}
\def\thesubsubsection{\arabic{subsubsection}.}
\def\appendix{\setcounter{section}{0}
 \def\thesection{Appendix \Alph{section}}
 \def\theequation{\Alph{section}.\arabic{equation}}}
\newcommand{\beq}{\begin{equation}}
\newcommand{\eeq}{\end{equation}}
\newcommand{\bea}{\begin{eqnarray}}
\newcommand{\eea}{\end{eqnarray}}
\newcommand{\beas}{\begin{eqnarray*}}
\newcommand{\eeas}{\end{eqnarray*}}
\newcommand{\defi}{\stackrel{\rm def}{=}}
\newcommand{\non}{\nonumber}
\def\de{\partial}
\def\si{\sigma}
\def\dim{\hbox{\rm dim}}
\def\sup{\hbox{\rm sup}}
\def\inf{\hbox{\rm inf}}
\def\Im{\hbox{\rm Im}}
\def\Re{\hbox{\rm Re}}
\def\Res{\hbox{\rm Res}}
\def\Max{\hbox{\rm Max}}
\def\infi{\infty}
\def\nrm{\parallel}
\def\de{\partial}
\def\om{\Omega}
\def\half{{1\over 2}}
\def\gs{{G(s)}}
\begin{titlepage}
\begin{center}
{\Large
Convergence of Scaled Delta Expansion: Anharmonic Oscillator}
\end{center}
\vspace{1ex}
\begin{center}
{\large
Riccardo Guida$^{1,2}$, Kenichi Konishi$^{1,2,\dagger}$ and Hiroshi
Suzuki$^{2,\ast}$}
\end{center}
\vspace{2ex}
\begin{center}
{\it $^{1}$ Dipartimento di Fisica -- Universit\`a di Genova\\
     Via Dodecaneso, 33 -- 16146 Genova (Italy)\\
     $^{2}$ Istituto Nazionale di Fisica Nucleare -- sez.~di Genova\\
     Via Dodecaneso, 33 -- 16146 Genova (Italy)\\
\vspace{1ex}
    e-mail addresses:
    decnet: 32655::x; internet: x@ge.infn.it\\
     x=konishi, guida}
\end{center}
\vfill
{\bf ABSTRACT:}

We prove that the linear delta expansion for energy eigenvalues of
the quantum mechanical anharmonic oscillator converges to the exact answer
if the order dependent trial frequency $\Omega$ is chosen to scale with
the order as $\Omega=CN^\gamma$; $1/3<\gamma<1/2$, $C>0$ as
$N\rightarrow\infty$. It converges also for $\gamma=1/3$, if
$C\geq\alpha_c g^{1/3}$, $\alpha_c\simeq 0.570875$, where $g$ is the
coupling constant in front of the operator $q^4/4$.
The extreme case with $\gamma=1/3$, $C=\alpha_cg^{1/3}$ corresponds to
the choice discussed earlier by Seznec and Zinn-Justin and,
more recently, by Duncan and Jones.
\vfill
\begin{flushleft}
GEF-Th-7/1994\hfill July 1994
\end{flushleft}
\vspace{2ex}
$^\dagger$ Address for September 1994--June 1995:
Univ.\ de Paris, Centre d'Orsay, France.\hfill\break
$^\ast$ On leave of absence from Department of Physics, Ibaraki University,
Mito 310, Japan (address after 1  September 1994).
\end{titlepage}
\section{Introduction}
\label{sec:intro}

A series of elegant papers in the seventies has explored and clarified various
aspects of the large order behavior in perturbation theory in quantum
mechanical and quantum field theoretic systems \cite{BW,Lip}. In some
cases this newly acquired knowledge was successfully used to obtain more
accurate results from the perturbative series, via, e.g., Borel summation
method. In other cases, especially in four-dimensional field theoretic models,
the question of what the {\em sum\/} of a badly-behaved perturbation
series means, remains unanswered. For a review, see \cite{GZ}. 

Recently, there has been a considerable interest in the so-called optimized
linear delta expansion [4--17]. In simple quantum mechanical models, the
method involves a rearrangement of the Hamiltonian as
\bea
   H&=&{p^2\over2}+V(q)
\\
   &=&H_0+\delta H'|_{\delta=1},
\eea
where
\begin{equation}
   H_0\equiv{p^2\over2}+{\Omega^2\over2}q^2,\quad
   H'\equiv V(q)-{\Omega^2\over2}q^2,
\end{equation}
where $\Omega$ is a trial frequency. The model is then expanded in
$\delta$ up to a given order $N$ according to the standard perturbation theory.
The exact result for any physical quantity should not depend on the trial
frequency $\Omega$ artificially introduced above, but the corresponding finite
order result $S_N$ does.

A proposal used often to fix $\Omega$ is to use
the ``principle of minimum sensitivity"~\cite{PMS}, i.e., to require that
$S_N$ be as little sensitive as possible to the variation of $\Omega$,
\beq
   {\partial S_N\over\partial\Omega}=0,
\eeq
which determines $\Omega$ and hence $S_N$ order by order. Another possible
criterion is the so-called ``fastest apparent convergence'' condition
(see \cite{DJ})
\begin{equation}
   S_N-S_{N-1}=0.
\label{fac}
\end{equation}
For the anharmonic oscillator, $V(q)=gq^4/4$, the optimized delta expansion
with both prescriptions gives surprisingly good results already at the lowest
nontrivial order, for any value of the coupling constant.

The method seems very powerful and, in our opinion deserves a careful study.
Already at first glance it displays several remarkable features.
First, the method uses the standard perturbative technique. This makes
the method potentially applicable to complicated systems such as field
theoretic models of fundamental interactions (Quantum Chromodynamics,
Quantum Electrodynamics, or the Weinberg--Salam theory), as well as to a wide
class of quantum mechanical systems.

Secondly, the method is nevertheless nonperturbative with respect to the usual
coupling constant since the order-dependent determination of $\Omega$
introduces a non-analytic dependence on it. The trial frequency $\Omega$,
adjusted so as to eliminate the higher-order disaster of the standard
perturbation theory, may be interpreted as a ``vacuum parameter," similar
to a certain field condensate or a physical mass parameter in the
effective Lagrangian method.

A particularly intriguing statement made by Duncan and Jones \cite{DJ} is that
the optimized delta expansion (apparently) cures the problem of large order
divergences even in non-Borel summable cases. If this were true in field
theoretic models it could shed some important light on the instanton
physics and vacuum structure of Quantum Chromodynamics.

Finally, the method can be regarded as a generalization of the standard
variational method, even though beyond the lowest order it involves no true
variational principle.

Despite these remarkable features, the theoretical basis of the success of
optimized delta expansion remains unclear. A key observation may be that the
method fails badly in a class of cases, where tunnelling effects are important
(e.g., for the low-lying energy eigenvalues of the quantum mechanical
double-well). If a method works in some cases but not in some other cases,
it should be possible to understand the reason for that, and through such
an analysis, to gain a better understanding of the mechanism underlying
the success and of the limit of its validity.

An important step forward has been made recently \cite{DJ}:
It was demonstrated that the optimized delta expansion converges to the
correct answer in the quantum mechanical anharmonic oscillator as well as
in the double well, at {\em finite temperatures}. At zero temperature
(hence for energy eigenvalues, Green functions, etc.), however, the
convergence proof does not apply. The proof was then extended \cite{BDJ}
in a zero dimensional model (an ordinary integral), to the logarithm of
the integral which is an analogue of the connected generating functional.

In this paper, we prove that the delta expansion for any energy eigenvalue of
the quantum mechanical anharmonic oscillator,
\beq
   H={p^2\over2}+{\omega^2\over2}q^2+{g\over4}q^4,
\label{aho}
\eeq
converges to the exact answer, if the order-dependent frequency $\Omega$
is chosen to scale with the order $N$ (asymptotically) as:
\beq
   x_N\equiv{\Omega_N\over\omega}=CN^\gamma,
\label{scaling}
\eeq
where either
\beq
   {1\over3}<\gamma<{1\over2},\quad C>0,
\label{index}
\eeq
or
\begin{equation}
   \gamma=1/3,\quad C\geq\alpha_c g^{1/3},\quad \alpha_c\simeq
   0.5708751028937741.
\label{newindex}
\end{equation}

In a zero-dimensional analogue model,
\beq
   Z(g,\omega)=\int_{-\infty}^\infty
   {dq\over\sqrt{2\pi}}\,
   \exp\left[-\left({1\over2}\omega^2q^2+{1\over4}gq^4\right)\right],
\eeq
a similar result holds (see \ref{sec:appa}) but with the scaling index in
a wider range
\beq
   {1\over4}<\gamma<{1\over2},
\eeq
or
\begin{equation}
   \gamma=1/4,\quad C\geq\alpha_cg^{1/4},
\end{equation}
where in this case $\alpha_c$ is given by Eq.~(\ref{alphazero}).

Our convergence proof however does not apply to the double well case
($\omega^2<0$).

Our work has some formal resemblance to the works by Buckley, Duncan,
Jones and Bender \cite{buck,DJ,BDJ} and has in fact been inspired by them,
but nevertheless differs considerably from theirs both in the method of
analysis and in the results found (our result for the extreme case with
$\alpha=\alpha_c$ however constitutes a generalization of their results).

The principle of minimum sensitivity or the fastest apparent convergence
criterion does not play any central role in the present work: the crucial
idea of what we call scaled delta expansion is to scale appropriately
the trial frequency with the order of expansion. In this respect our
philosophy is close to a similar idea expressed by Bender, Duncan and Jones
\cite{BDJ}, but we carry it to an extreme (relying solely on it),
whereas their particular scaling behavior was suggested by the
optimization procedure for $Z(g,\omega)$.

The ideas underlying our proof can be explained as follows. For dimensional
reasons any energy eigenvalue of the anharmonic oscillator has the form,
\beq
   E(g,\omega)=\omega\widetilde E\left({g\over\omega^3}\right).
\eeq

The standard (asymptotic) perturbation series then reads formally
\beq
   E^{(pert)}=\omega\sum_{n=0}^{\infty}c_n\left({g\over\omega^3}\right)^n,
\label{per}
\eeq
where the coefficients $c_n$ have the known large order behavior at
$N\rightarrow\infty$ \cite{BW,GZ},
\beq
   c_n\sim
   -{\sqrt{6}\,12^K\over\pi^{3/2}K!}\left(-{3\over 4}\right)^n
   \Gamma(n+K+1/2).
\eeq

The delta expansion for Eq.~(\ref{aho}), i.e., a perturbative expansion in
\beq
   H'={\omega^2-\Omega^2\over2}q^2+{g\over4}q^4,
\label{Hprime}
\eeq
is equivalent to a substitution in Eq.~(\ref{per}) \cite{GZ,Neveu}
\bea
   &&\omega\rightarrow
   \Omega\left(1+\delta\,{\omega^2-\Omega^2\over\Omega^2}\right)^{1/2}
   =\omega x[1-\delta\beta(x)]^{1/2},
\nonumber
\\
   &&g\rightarrow\delta\cdot g,
\label{sub}
\eea
where
\begin{equation}
   x\equiv{\Omega\over\omega},\quad
   \beta(x)\equiv1-{1\over x^2},
\end{equation}
followed by the expansion in $\delta$. A subsequent rearrangement of
the series yields
\bea
   &&E^{(\delta)}=\omega\lim_{N\rightarrow\infty}S_N(x),
\nonumber
\\
   &&S_N(x)=\sum_{n=0}^Nc_n\left({g\over\omega^3}\right)^nA_n^{(N)}(x),
\nonumber
\\
   &&A_n^{(N)}(x)\equiv{1\over x^{3n-1}}\sum_{k=0}^{N-n}\beta(x)^k
     {\Gamma(3n/2+k-1/2)\over\Gamma(3n/2-1/2)\Gamma(k+1)}.
\label{delta}
\eea

A parenthetical remark, useful for numerical analysis, is that the delta
expansion for the {\em double well\/} case [$\omega^2<0$ in Eq.~(\ref{aho})]
is given exactly by Eq.~(\ref{delta}) (in particular with the {\em same\/}
coefficients $c_n$ of the anharmonic oscillator), the only modification being
\bea
   &&\omega\rightarrow\sqrt{-\omega^2},
\nonumber
\\
   &&x\rightarrow x=\Omega/\sqrt{-\omega^2},
\nonumber
\\
   &&\beta(x)\rightarrow\beta_{\rm DW}(x)=1+1/x^2.
\nonumber
\label{DWmod}
\eea
This assertion may at first sight appear puzzling but is actually self-evident
since the sign change of $\omega^2$ causes only trivial changes in the
coefficients order by order of the expansion in $H'$ [see Eq.~(\ref{Hprime})].

Coming back to the anharmonic oscillator, $S_N(x)$ has the form of the standard
perturbation series with its coefficients modified by the
$\Omega$-dependent factors $A_n^{(N)}(x)$. Except for $A_0^{(N)}(x)$ they
are positive definite numbers such that
\beq
   0<A_n^{(N)}(x)<1\quad{\rm for}\quad n\ge1,
\eeq
(we assume $x>1$ throughout this paper). All of them [including
$A_0^{(N)}(x)$] converges to unity as $N\rightarrow\infty$ if $x$ is
kept fixed, while they all vanish [except for $A_0^{(N)}(x)$ which
diverges] if $x\rightarrow\infty$ with $N$ fixed. $S_N(x)$ can thus be
regarded as a sort of regularized (or stabilized) perturbation series.
The idea is then to play with the rates at which $N$ and $x$
are sent {\em simultaneously\/} to infinity so that the convergence is
ensured.

The convergence at the upper side of the sum ($n=N$) requires that
\beq
   c_NA_N^{(N)}(x)\sim N!/x^{3N-1}\rightarrow0,
\eeq
this suggests $x$ be scaled as $x\propto N^\gamma$, $\gamma>1/3$. On the
other hand, to guarantee the convergence of $A_0^{(N)}(x)$ in the large
$N$ limit it is necessary that
\beq
   \beta(x)^N = (1-1/x^2)^N \rightarrow0,
\eeq
this implies that the scaling index $\gamma$ to be less than $1/2$.
The detailed study of the next section fully confirms these expectations.

This paper is organized as follows. In Section~\ref{sec:proof}
we prove rigorously that the sequence $\{S_N\}$ converges to the exact answer
as $N$, $x\rightarrow\infty$ according to the scaling in Eqs.~(\ref{scaling})
and (\ref{index}). The convergence proof for the case $\gamma=1/3$ is
given in Section~\ref{sec:oneoverthree}. The uniformity of convergence,
as well as the manner in which $S_N$ diverges when the scaling index is
outside the convergence domain are discussed in Section~\ref{sec:outside}.
These properties are compared to the numerical calculations of the delta
expansion carried out to high orders ($N\sim 100$) in
Section~\ref{sec:numerical}. In Section~\ref{sec:orderdep} we reinterpret
our result in terms of Seznec--Zinn-Justin's order dependent mappings.

We conclude the paper by discussing several features of the scaled
delta expansion and their possible generalizations in
Section~\ref{sec:discussion}. \ref{sec:appa} gives a convergence proof of the
scaled delta expansion in the zero dimensional (ordinary integral) case;
\ref{sec:appb} discusses the properties of the roots of an equation used
in Section~\ref{sec:proof}.

\section{Scaled delta expansion for the anharmonic oscillator}
\label{sec:mainpart}

In this section the scaled delta expansion for the anharmonic oscillator is
discussed in detail.
\subsection{Proof of convergence for $1/3<\gamma<1/2$}
\label{sec:proof}

\myfigure{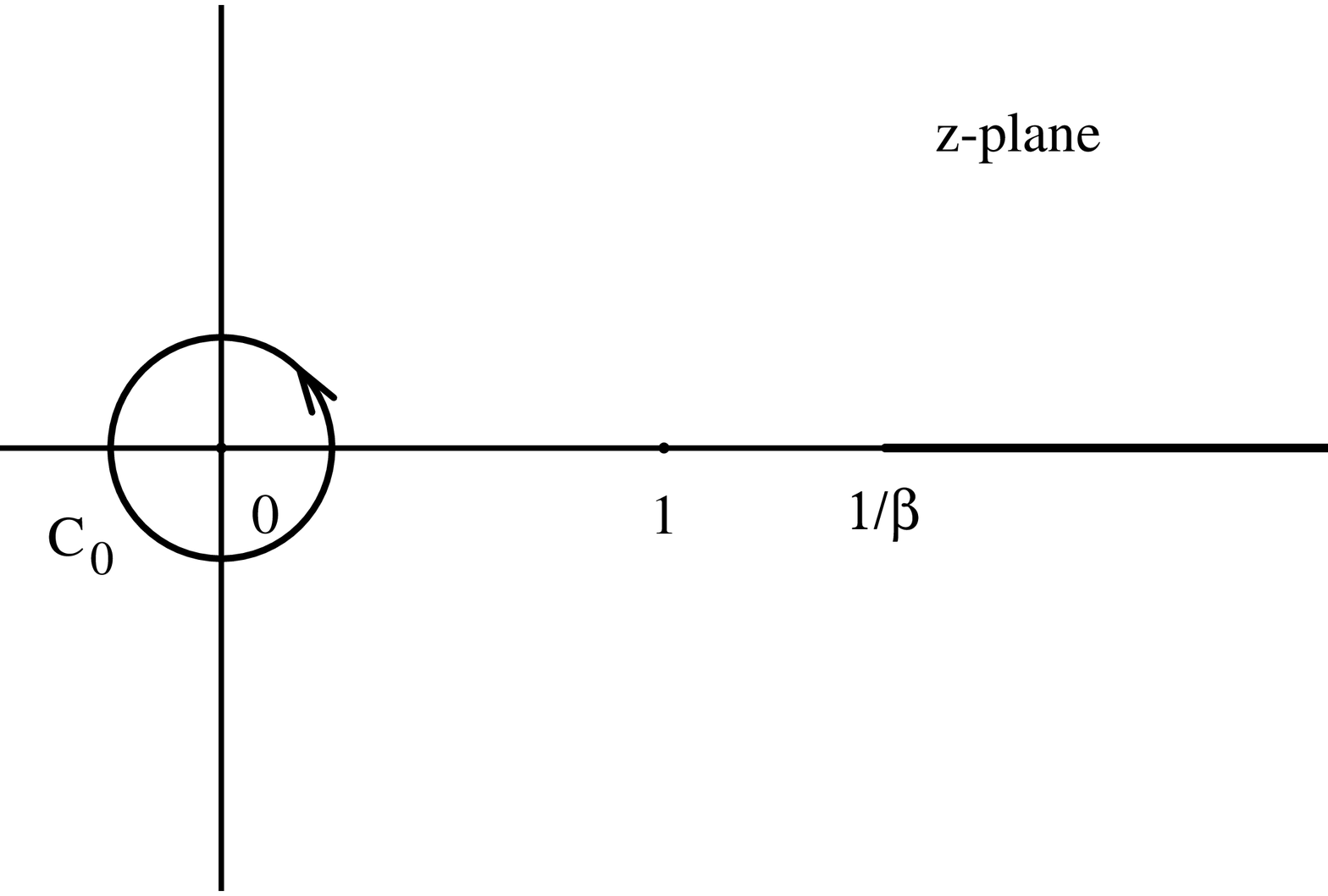}{0.5}{$C_0$ contour in the complex $z$-plane.}
First we prove rigorously that the scaled delta expansion for the anharmonic
oscillator converges to the exact answer for $K$-th excited energy level
($K=0,1,\cdots$), if the scaling of $x_N$ is chosen in the range
Eq.~(\ref{index}). Essential ingredients for the proof are:
\begin{description}
\item{1)}
that an energy eigenvalue of anharmonic oscillator satisfies a once-subtracted
dispersion relation \cite{LM},
\bea
   E(g,\omega)&=&
   c_0\omega+{g\over\pi}\int_{-\infty}^0dg'\,{\Im\,E(g')\over g'(g'-g)}
\nonumber
\\
   &=&c_0\omega+{g\over\pi\omega^2}\int_0^{\infty}d\lambda\,
   {\Im\,\widetilde E(\lambda)\over\lambda(\lambda+g/\omega^3)},
\label{disp}
\eea
where $c_0=K+1/2$ and in the second line dimensionless energy and coupling
constant
\beq
   \widetilde E(\lambda)\equiv E(g',\omega)/\omega,\quad
   \lambda\equiv-g'/\omega^3,
\eeq
have been introduced. This relation follows from the analytic structure
of the energy in the complex coupling plane, and from the known large
$\lambda$ behavior $\widetilde E(\lambda)\sim\lambda^{1/3}$.
By expanding Eq.~(\ref{disp}) in powers of $g/\omega^3$ one finds an
expression for the perturbative coefficients $c_n$ in terms of an integral
involving $\Im\,\widetilde E(\lambda)$ which was useful in the determination
of their large order behavior~\cite{GZ};

\item{2)}
the behavior of $\Im\,\widetilde E(\lambda)$ at small negative coupling
constant (small positive $\lambda$) determined by the tunnelling factor
\cite{BW},
\beq
   \Im\,\widetilde E(\lambda)={4^{2K+1}\over\sqrt{2\pi}K!}
   \lambda^{-K-1/2}\exp\left(-{4\over3\lambda}\right)
   \left[1 + O(\lambda)\right],
\label{smallg}
\eeq
for the $K$-th energy level; and

\item{3)}
the positivity and boundedness of $\Im\,\widetilde E(\lambda)/\lambda^2$
\cite{LM},
\beq
   \infty>\Im\,\widetilde E(\lambda)/\lambda^2>0,
   \quad{\rm for}\quad\lambda>0.
\eeq
[This is a minor technical assumption. For the following proof,
$\Im\,\widetilde E(\lambda)/\lambda^2\in L^1[0,\infty]$ is sufficient.]
\end{description}

\myfigure{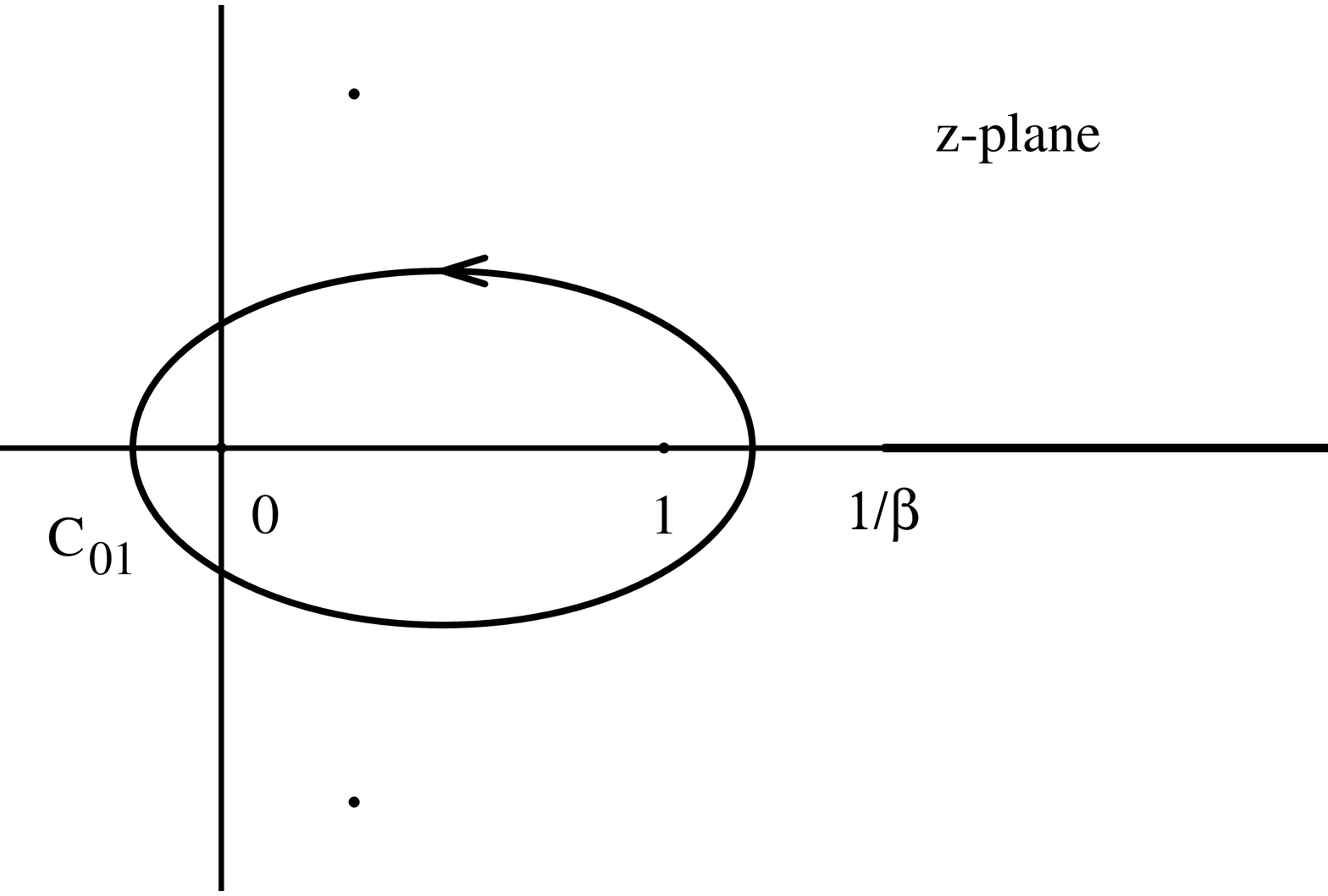}{0.4}{$C_{01}$ contour in the complex $z$-plane.}
From now on $\omega$ will be set to unity, $\omega=1$.
The $N$-th order delta expansion approximant for $E$, $S_N$, can be readily
constructed by the substitution rule Eq.~(\ref{sub}) applied to
Eq.~(\ref{disp}), followed by the Taylor expansion on $\delta$ up to
$\delta^N$ ($\delta=1$ at the end). This can be expressed compactly with
the use of Cauchy's formula as:
\bea
   &&S_N(x)=S_{0N}(x)+S_{1N}(x),
\nonumber
\\
   &&S_{0N}(x)\equiv
   c_0x\oint_{C_0}{dz\over2\pi i}{1-1/z^{N+1}\over z-1}(1-\beta z)^{1/2},
\nonumber
\\
   &&S_{1N}(x)\equiv
   {g\over\pi x^2\beta}\int_0^{\infty}{d\lambda\over\lambda^2}\,
    \Im\,\widetilde E(\lambda)\oint_{C_0}{dz\over2\pi i}
    {1-1/z^{N+1}\over z-1}F(\beta z,s),
\label{S_N}
\eea
where
\beq
   F(w,s)\equiv{w(1-w)^{1/2}\over(1-w)^{3/2}+sw},
\label{F}
\eeq
and
\beq
   s\equiv{g\over x^3\beta\lambda},\quad\beta(x)\equiv1-{1\over x^2},
\label{sbeta}
\eeq
and the contour $C_0$ is a small circle around the origin (Fig.~1).
Because the integrand has no pole at $z=1$, 
the integration contour may be deformed as in Fig.~2.\footnote{Note that this
deformation does not encounter any obstruction due to the cut starting at
$z=1/\beta>1$ or due to the poles of $F(\beta z,s)$: see below.}
 Noting that 
the first term of the integrand ($1$ in the numerator) gives the exact
energy in Eq.~(\ref{disp}), we get the following expression for the remainder
\bea
   &&R_N(x)\equiv E(g,\omega)-S_N(x)=R_{0N}(x)+R_{1N}(x),
\label{remainder}
\\
   &&R_{0N}(x)\equiv
     c_0x\oint_{C_{01}}{dz\over2\pi i}{(1-\beta z)^{1/2}\over z^{N+1}(z-1)},
\\
   &&R_{1N}(x)\equiv{g\over\beta x^2}\int_0^{\infty}{ds\over s}\,G(s)
     \oint_{C_{01}}{dz\over2\pi i}{F(\beta z,s)\over z^{N+1}(z-1)},
\label{R1}
\eea
where the contour $C_{01}$ now encircles $z=0$ and $z=1$ but no other
singularities (see Fig.~2). Note that in Eq.~(\ref{R1}) the integration
variable has been changed from $\lambda$ to $s$ in Eq.~(\ref{sbeta}) and
\beq
   G(s)\equiv{\Im\,\widetilde E(\lambda)\over\pi\lambda}
   \biggr|_{\lambda=g/(x^3\beta s)}.
\label{gfunc}
\eeq

\myfigure{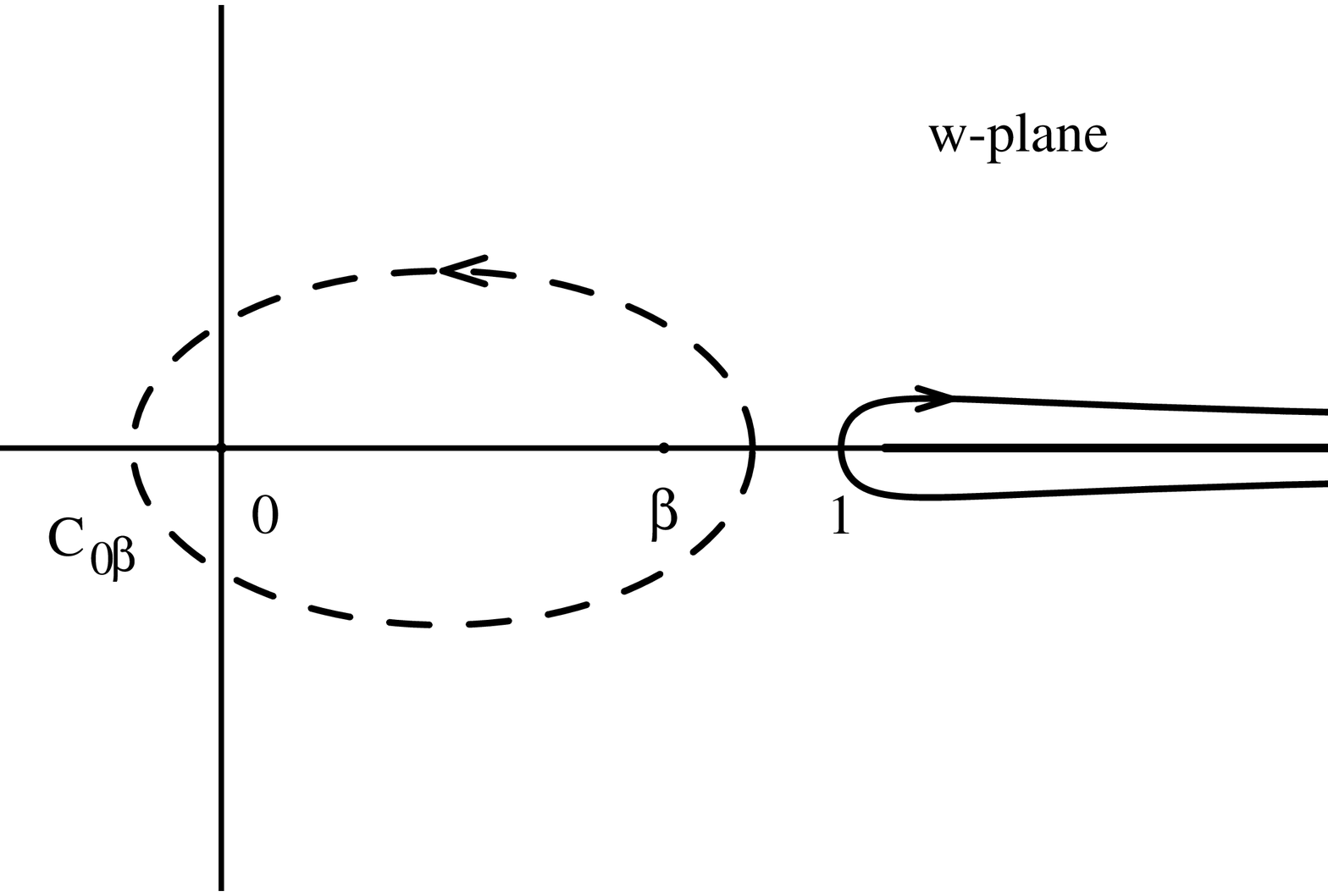}{0.4}
{
Deformation of $C_{0\beta}$ contour ($w$-plane) into cut contribution
for $R_{0N}$.
}

Let us consider $R_{0N}(x)$ and $R_{1N}(x)$ separately. First consider
\beq
   R_{0N}(x)=c_0x\beta^{N+1}\oint_{C_{0\beta}}{dw\over2\pi i}
   {(1-w)^{1/2}\over w^{N+1}(w-\beta)},
\eeq
where the integration variable has been changed to $ w=\beta z$. The contour
can now be deformed further to wrap around the cut on the real positive axis
$[1,\infty)$ (see Fig.~3).
 The circle at infinity does not give any
contribution. Introducing a real variable $u$ by $w =1+u\pm i\epsilon$,
one finds
\beq
   R_{0N}(x)=-{c_0x\beta^{N+1}\over\pi}
   \int_0^\infty du\,{u^{1/2}\over(u+1-\beta)(1+u)^{N+1}}.
\label{R0N}
\eeq
Now
\beq
   \left|R_{0N}(x)\right|<{c_0x\beta^{N+1}\over\pi}
   \int_0^\infty du\,{u^{-1/2}\over(1+u)^{N+1}}
   ={c_0x\beta^{N+1}\over\pi}B(1/2,N+1/2),
\label{bound11}
\eeq
where $B(1/2,N+1/2)$ is the Euler's Beta function. Using the Stirling's
formula we see:
\beq
   \left|R_{0N}(x)\right|<{c_0x\beta^{N+1}\over\sqrt{\pi}N^{1/2}}
   \left[1+O(1/N)\right].
\label{bound1}
\eeq

Next consider
\beq
   R_{1N}(x)={g\beta^N\over x^2}\int_0^{\infty}{ds\over s}\,G(s)
   \oint_{C_{0\beta}}{dw\over2\pi i}{F(w,s)\over w^{N+1}(w-\beta)},
\eeq
where $w=\beta z$. In deforming further the integration contour the poles of
the function $F(w,s)$ [see Eq.~(\ref{F})] must be taken into account this
time, as well as the cut. See Fig.~4.
 Accordingly $R_{1N}(x)$ can be split as
\bea
   R_{1N}(x)&=&{g\beta^{N}\over x^2}\int_0^{\infty}{ds\over s}\,G(s)
   (\sum_{i={\rm poles}}{\cal R}_i+{\cal C})
\nonumber
\\
   &\equiv&\sum_{i={\rm poles}}R_{1N}^{(i)}(x)+R_{1N}^{\rm (cut)}(x)
\label{split}
\eea
where
\beq
   {\cal R}_i=-\Res\left[{F(w,s)\over w^{N+1}(w-\beta)}\right]_{w=w_i},
\label{poles}
\eeq
is the residue of the $i$-th pole $w_i$, i.e., at the $i$-th root of the
equation
\beq
   (1-w)^{3/2}+sw=0,
\label{roots}
\eeq
and
\beq
   {\cal C}=-{1\over\pi}\int_0^\infty du\,
   {su^{1/2}\over(1+u)^{N-1}(1-\beta+u)\left[u^3 + s^2(1+u)^2\right]}.
\label{cut}
\eeq

\myfigure{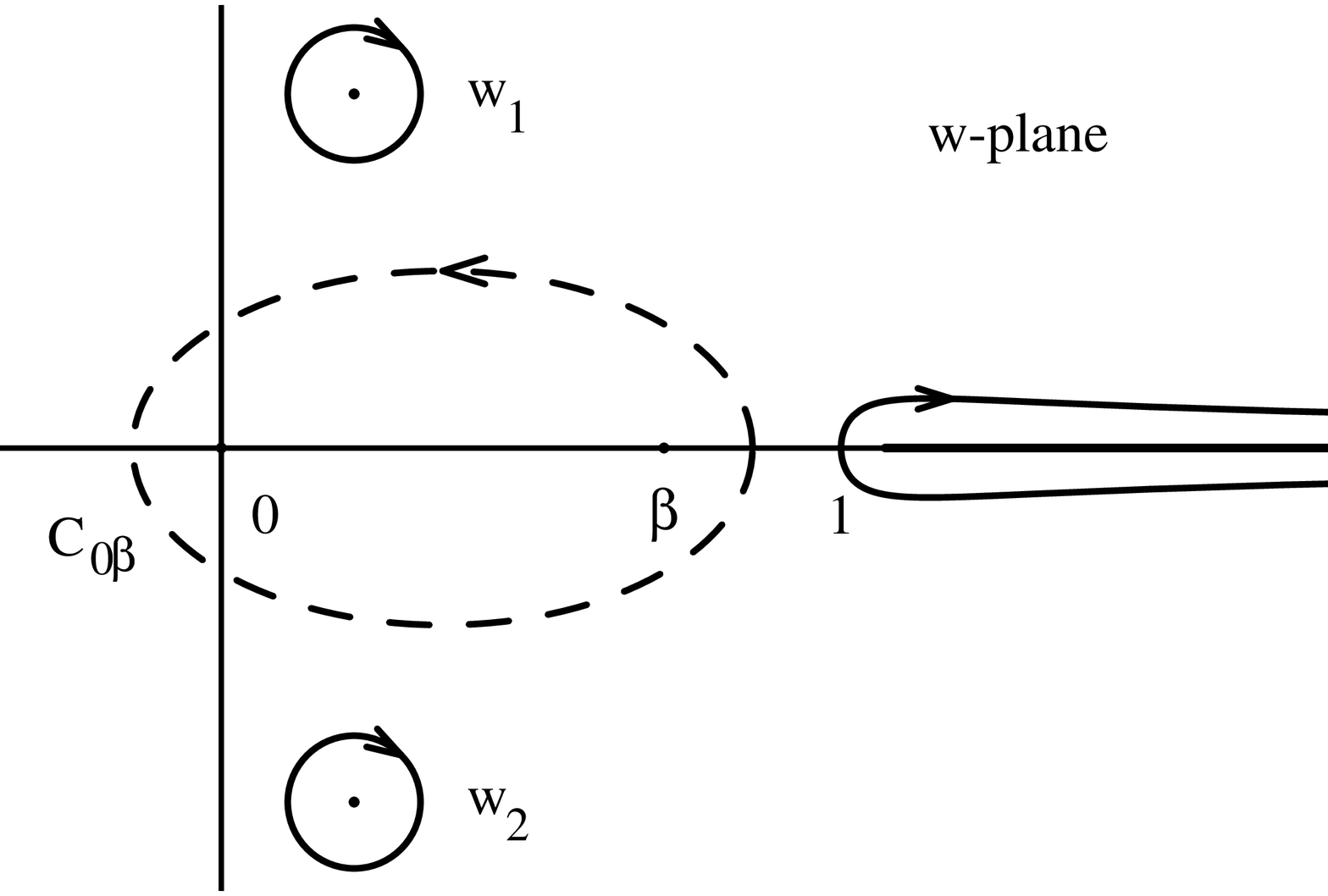}{0.4}
{
Deformation of $C_{0\beta}$ contour ($w$-plane) into cut and poles
contributions for $R_{1N}$.
}
The bound on $R_{1N}^{\rm (cut)}(x)$ can be set easily as follows. Obviously
($1-\beta=1/x^2$),
\begin{equation}
   |{\cal C}|<{x^2\over\pi}I(s),\quad
   I(s)\equiv\int_0^{\infty}du\,{su^{1/2}\over u^3+s^2(1+u)^2}.
\label{cbound}
\end{equation}
But $I(s)$ is continuous and behaves as 
\begin{equation}
   I(s)\rightarrow{\pi\over3}\quad{\rm as}\quad s\rightarrow0,
   \quad{\rm and}\quad
   I(s)\sim{\pi\over2s}\quad{\rm as}\quad s\rightarrow\infty,
\end{equation}
so that it is bounded in $[0,\infty)$. The contribution of the cut to
$R_{1N}(x)$ is thus bounded by [see Eq.~(\ref{split}), Eq.~(\ref{cbound})]
\beq
   \left|R_{1N}^{\rm (cut)}(x)\right|<{\rm const.}\,c_1g\beta^{N},
\label{bound2}
\eeq
where we have used
\begin{equation}
   \int_0^\infty{ds\over s}\,G(s)=
   {1\over\pi}\int_0^\infty d\lambda\,
   {\Im\,\widetilde E(\lambda)\over\lambda^2}=
   c_1={3\over16}(2K^2+2K+1),
\end{equation}
($c_1$ is the first coefficient of the standard perturbative series).

To set the bound on $\sum_{i={\rm poles}} R_{1N}^{(i)}(x)$ requires a
little more work. The properties of the roots of Eq.~(\ref{roots}) which
lie on the first Riemann sheet (they are the only ones that interest us)
are studied in \ref{sec:appb}.
There are two such roots $w_1$ and $w_2$ which move around on the first
Riemann sheet as shown in Fig.~5 as $s$ varies from $0$ to $\infty$. At $s=0$,
they are coincident at $w_1=w_2=1$; when $0<s<s_c\equiv3\sqrt{3}/2$, they
are complex conjugate to each other and lie on the two branches of the
apple-like curve in Fig.~5; at $s=s_c$ they coalesce again at $w_1=w_2=-2$.
For larger values of $s$, $s>s_c$, they are real and stay on the negative
real axis: $-\infty<w_1<-2$, and $-2<w_2<0$. Note that they keep always
outside the integration contour $C_{0\beta}$ of Fig. 4.

Near its $i$-th pole $w\simeq w_i$ the function $F(w,s)$ behaves as
\beq
   F(w,s)\simeq-{2w_i^2\over w_i+2}{1\over w-w_i},
\eeq
[where Eq.~(\ref{roots}) has been used] hence
\beq
   {\cal R}_i={f(w_i)\over w_i+2},
\eeq
with
\beq
   f(w)\equiv{2\over w^{N-1}(w-\beta)}.
\eeq

A little obstacle arises in setting the bound on
$\sum_{i={\rm poles}}R_{1N}^{(i)}(x)$ due to the large residues near
$s=s_c$ where $ w_i\simeq-2$. Actually (see \ref{sec:appb}), in the vicinity
of $-2$ the two roots are at symmetric positions with respect to $-2$, i.e.,
\beq
   w_{1,2}=-2\mp4\sqrt{\epsilon}/\sqrt[4]{3},\quad s=3\sqrt{3}/2+\epsilon,
\eeq
so that these two large residues cancel in the sum. At this point it is
convenient to split the $s$ integration range into three parts:
(I) $0\le s<s_0$, (II) $s_0\le s<s_c$, (III) $s_c\le s<\infty$, where
$s_0=2\sqrt{2/3}$ is the value of $s$ at which $w_1$ is purely imaginary
$w_1(s_0)=\sqrt{3}i$ (see Fig.~5). Any other choice of $s_0<s_c$ would work.

\myfigure{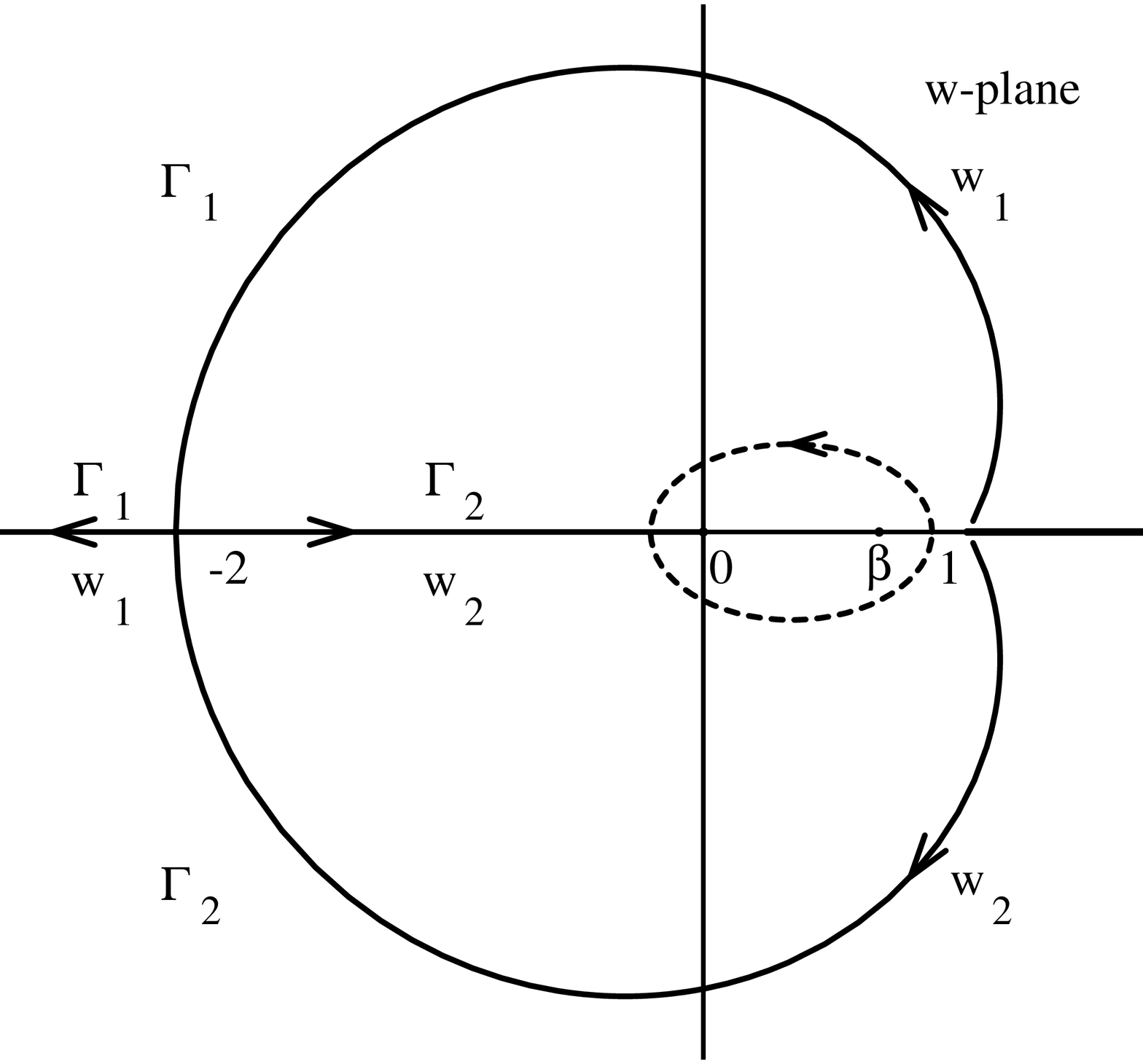}{0.45}
{
Behavior of the poles $w_1$ and $w_2$ in the complex $w$-plane
as the parameter $s$ varies.
}                      

\noindent
(I) $0\le s<s_0$.

In this region, two poles are complex conjugate to each other,
$w_2 = w_1^\ast$, therefore
\beq
   \left|\sum_{i=1,2}R_{1N}^{(i)}(x)\right|_{\rm (I)}
   \le2\left|R_{1N}^{(1)}(x)\right|_{\rm (I)}.
\label{respoles}
\eeq
For the right hand side, we find
\bea
   \left|R_{1N}^{(1)}(x)\right|_{\rm (I)}
   &\le&{g\beta^{N}\over x^2}\int_0^{s_0}{ds\over s}\,G(s)
   \left|{f(w_1)\over w_1+2}\right|
\nonumber
\\
   &<&g\beta^{N}\int_0^{s_0}{ds\over s}\,G(s)
\nonumber
\\
   &<&c_1g\beta^{N},
\label{bound7}
\eea
where inequalities
\beq
   |w_1|\ge1,\quad|w_1-\beta|\ge1-\beta={1\over x^2},\quad|w_1+2|>2,
\eeq
(see Fig.~5) have been used. In deriving the last line of Eq.~(\ref{bound7}),
the integration region of $s$ has been extended to $[0,\infty)$.
 
For regions (II) and (III), the difficulty of infinite residues at
$s=s_c$ can be overcome by rewriting the pole contribution to
Eq.~(\ref{split}) as:
\bea
   &&\left[\sum_{i=1,2}R_{1N}^{(i)}(x)\right]_{{\rm (II)}+{\rm (III)}}
\nonumber
\\
   &&=\sum_{i=1,2}{g\beta^N\over x^2}
   \int_{s_0}^\infty{ds\over s}\,G(s){f(w_i)-f(-2)\over w_i+2}
   +{g\beta^N\over x^2}
   \int_{s_0}^\infty{ds\over s}\,G(s)f(-2)\sum_{i=1,2}{1\over w_i+2}
\nonumber
\\
   &&\equiv\sum_{i=1,2}\widetilde R_{1N}^{(i)}(x)+\widetilde R_{1N}^{(0)}(x),
\label{trick}
\eea
with obvious definitions for $\widetilde R_{1N}^{(i)}(x)$ and
$\widetilde R_{1N}^{(0)}(x)$. Bounds on each of $\widetilde R_{1N}^{(i)}(x)$
and $\widetilde R_{1N}^{(0)}(x)$ can now be set without difficulty.

First consider $\widetilde R_{1N}^{(0)}(x)$. The function $h(s)\equiv
\sum_{i=1,2}1/[w_i(s)+2]$ is continuous for $s$ in $[0,\infty)$
(including $s=s_c$) and $\lim_{s\rightarrow\infty}h(s)=1/2$, hence
is bounded. Therefore
\beq
   \left|\widetilde R_{1N}^{(0)}(x)\right|
   \le{\rm const.}\,{g\beta^N\over2^Nx^2}\int_{s_0}^\infty{ds\over s}\,G(s)
   <{\rm const.}\,{c_1g\beta^N\over2^Nx^2}.
\label{bound3}
\eeq

Next consider $\widetilde R_{1N}^{(i)}(x)$ ($i=1$, $2$) each of which is
well-defined. We first note that
\beq
   |f(w)-f(-2)|
   =\left|\int_\Gamma dw'\,f'(w')\right|
   \le\max_\Gamma|f'(w)|\cdot L_\Gamma,
\label{arc}
\eeq
where $\Gamma$ is a path connecting the points $-2$ and $w$ and $L_\Gamma$
is its length. Let us choose as $\Gamma_i$ the trajectory $w=w_i(s')$
traced by the $i$-th root as $s'$ moves away from $s_c =3\sqrt{3}/2$
to the current value of $s$. Thus we define:
\bea
   \Gamma_i&\equiv&
   \left\{w\,\bigg|\,w=w_i(s'), s_c\leq s'\leq s\right\},\quad s\geq s_c,
\nonumber
\\
   \Gamma_i&\equiv&
   \left\{w\,\bigg|\,w=w_i(s'), s\leq s'\leq s_c\right\},\quad s<s_c
\eea
(see Fig.~5). $L_{\Gamma_i}$  then satisfies
\bea
   L_{\Gamma_i}(s)&=&|w_i(s)+2|,\quad s\geq s_c,
\nonumber
\\
   L_{\Gamma_i}(s)&\le&A'|w_i(s)+2|,\quad s< s_c,
\eea
where
\beq
   A'\equiv\max_{s\in[s_0,s_c]}{L_{\Gamma_i}(s)\over|w_i(s)+2|}
\eeq
is some finite constant  independent on $s$. It follows then that
\beq
   \left|{f(w_i)-f(-2)\over w_i+2}\right|
   \le{\rm const.}\,\max_{\Gamma_i}|f'(w)|
\label{max}
\eeq
in all cases. We use this relation and
\beq
   f'(w)=-f(w)\left({N-1\over w}+{1\over w-\beta}\right)
\label{fprimo}
\eeq
to get the bounds for $\widetilde R_{1N}^{(i)}(x)$ ($i=1$, $2$) in the
regions (II) and (III).

\noindent
(II) $s_0\leq s<s_c$.

In this region $w_1$ and $w_2$ are still complex conjugate, so that it
suffices to bound $\widetilde R_{1N}^{(1)}(x)$ [see Eq.~(\ref{respoles})].
Using the fact that $|w_1|$ is an increasing function of $s$
(see \ref{sec:appb}) we get:
\beq
   |w_1|\ge\sqrt{3},\quad|w_1-\beta|>\sqrt{3},
\eeq
thus from Eq.~(\ref{fprimo})
\beq
   \max_{\Gamma_1}|f'|<{2(N+1)\over3^{(N+1)/2}}.
\eeq
Use of Eq.~(\ref{max}) and Eq.~(\ref{trick}) then yields
\bea
   \left|{\widetilde R}_{1N}^{(1)}(x)\right|_{(II)}
   &<&{\rm const.}\,{gN\beta^N\over3^{N/2}x^2}\int_{s_0}^{s_c}{ds\over s}\,
   G(s)
\nonumber
\\
   &<&{\rm const.}\,{c_1gN\beta^N\over3^{N/2}x^2}.
\label{bound6}
\eea

\noindent
(III) $s_c\le s<\infty$.

In the region, a useful fact is that
\beq
   |f(w)|={2\over|w|^{N-1}|w-\beta|},
\eeq
is a monotonically increasing function of $w$ for negative $w$. For the
smaller root, $w_1\le-2$, this implies that $|f(w_1)|\le|f(-2)|$, hence
\beq
   \max_{\Gamma_1}|f'(w)|<|f(-2)|\left({N-1\over2}+{1\over2}\right)
   <{N\over2^N}.
\label{fprime}
\eeq
It follows that
\bea
   \left|\widetilde R_{1N}^{(1)}(x)\right|_{\rm (III)}
   &<&{gN\beta^N\over2^Nx^2}\int_{s_c}^\infty{ds\over s}\,G(s)
\nonumber
\\
   &<&{c_1gN\beta^N\over2^Nx^2}.
\label{bound4}
\eea

For the second root $w_2$ ($-2\le w_2<0$) we instead use the fact that
\beq
   w_2(s)=-{1\over s}[1-w_2(s)]^{3/2}\le-{1\over s}.
\eeq
Then
\beq
   |f(w_2)|\le|f(-1/s)|={2s^{N-1}\over\beta+1/s},
\eeq
yielding
\bea
   \max_{\Gamma_2}|f'(w)|&<&|f(-1/s)|\left[s(N-1)+{1\over\beta}\right]
\nonumber
\\
   &<&{2s^N\over\beta}\left(N+{1\over\beta s}\right).
\nonumber
\eea
This leads to an inequality:
\bea
   \left|\widetilde R_{1N}^{(2)}(x)\right|_{\rm (III)}
   &<&{2g\beta^{N-1}\over x^2}\int_{s_c}^\infty{ds\over s}\,
      s^NG(s)\left(N+{1\over\beta s}\right)
\nonumber
\\
   &\equiv&{2g\beta^{N-1}\over x^2}\left(NI_N+{1\over\beta}I_{N-1}\right),
\label{R2tilde}
\eea
where
\beq
   I_N\equiv\int_{s_c}^\infty{ds\over s}\,s^NG(s)
\eeq
and function $G(s)$ is defined in Eq.~(\ref{gfunc}). The contribution from
large $s$ is potentially dangerous. Fortunately, for $s\ge s_c$ and at
large $x$ (as we are interested in the limit $x\rightarrow\infty$)
$\lambda=g/(x^3\beta s)$ is always small, so the asymptotic estimate for
$\Im\,\widetilde E(\lambda)$ of Eq.~(\ref{smallg}) can be used:
\bea
   I_N&=&
   {4^{2K+1}\over\sqrt{2}\pi^{3/2}K!}\left({x^3\beta\over g}\right)^{K+3/2}
   \int_{s_c}^\infty ds\,s^{N+K+1/2}\exp\left(-{4x^3\beta s\over3g}\right)
   \left[1+O(g/(x^3\beta s))\right]
\nonumber
\\
    &<&{4^{2K+1}\over\sqrt{2}\pi^{3/2}K!}\left({3\over4}\right)^{N+K+3/2}
    \left({g\over x^3\beta}\right)^N\Gamma(N+K+3/2)\left[1+O(1/N)\right],
\eea
where $K$ is the energy level. Substituting this into Eq.~(\ref{R2tilde}) and
using Stirling's formula, we find
\beq
   \left|\widetilde R_{1N}^{(2)}(x)\right|_{\rm (III)}
   <{3\sqrt{3}\,12^KgN^{K+1}\over\pi K!x^2\beta}\left(N+{4x^3\over3gN}\right)
   \left(3gN\over4ex^3\right)^N\left[1+O(1/N)\right].
\label{bound5}
\eeq

We are now in a position to set an upper bound to the full remainder $R_N$
by using Eqs.~(\ref{remainder}), (\ref{split}), (\ref{respoles}) and
(\ref{trick}):
\begin{equation}
   \left|R_N\right|
   \le
   \left|R_{0N}\right|+\left|R_{1N}^{\rm (cut)}\right|
   +2\left|R_{1N}^{(1)}\right|_{\rm (I)}+\left|\widetilde R_{1N}^{(0)}\right|
   +2\left|\widetilde R_{1N}^{(1)}\right|_{\rm (II)}
   +\left|\widetilde R_{1N}^{(1)}\right|_{\rm (III)}
   +\left|\widetilde R_{1N}^{(2)}\right|_{\rm (III)},
\label{upperbound}
\end{equation}
with individual bounds given in 
Eqs.~(\ref{bound1}), (\ref{bound2}), (\ref{bound7}),
(\ref{bound3}), (\ref{bound6}), (\ref{bound4}) and (\ref{bound5}),
\bea
   &&\left|R_{0N}(x)\right|
     <{c_0x\beta^{N+1}\over\sqrt{\pi}N^{1/2}}\left[1+O(1/N)\right],
\nonumber
\\
   &&\left|R_{1N}^{\rm (cut)}(x)\right|<{\rm const.}\,c_1g\beta^N,
\nonumber
\\
   &&\left|R_{1N}^{(1)}(x)\right|_{\rm (I)}<c_1g\beta^{N},
\nonumber
\\
   &&\left|\widetilde R_{1N}^{(0)}(x)\right|<{\rm const.}\,
     {c_1g\beta^N\over2^Nx^2},
\label{summary}
\\
   &&\left|\widetilde R_{1N}^{(1)}(x)\right|_{\rm (II)}
     <{\rm const.}\,{c_1gN\beta^N\over3^{N/2}x^2},
\nonumber
\\
   &&\left|\widetilde R_{1N}^{(1)}(x)\right|_{\rm (III)}
     <{c_1gN\beta^N\over2^Nx^2},
\nonumber
\\
   &&\left|\widetilde R_{1N}^{(2)}(x)\right|_{\rm (III)}
     <{3\sqrt{3}\,12^KgN^{K+1}\over\pi K!x^2\beta}
      \left(N+{4x^3\over3gN}\right)
      \left(3gN\over4ex^3\right)^N\left[1+O(1/N)\right].
\nonumber
\eea

If the trial frequency $x=\Omega/\omega$ is scaled as
\beq
   x=CN^\gamma
\eeq
and the limit $N\rightarrow \infty$ is taken at fixed
$K$ and  $g$, then
\beq
   \left|R_{0N}\right|,\,\left|R_{1N}^{\rm (cut)}\right|,\,
   \left|R_{1N}^{(1)}\right|_{\rm (I)},\,
   \left|\widetilde R_{1N}^{(0)}\right|,\,
   \left|\widetilde R_{1N}^{(1)}\right|_{\rm (II)},\,
   \left|\widetilde R_{1N}^{(1)}\right|_{\rm (III)}
   \rightarrow 0,
\eeq
if $0<\gamma<1/2$, while
\beq
   \left|\widetilde R_{1N}^{(2)}\right|_{\rm (III)}\rightarrow 0,
\eeq
if $\gamma>1/3$. This completes our proof for $1/3<\gamma<1/2$.

Up to now no hypothesis was made on the constant $C$, which could be
chosen to depend on $g$ and $K$. Actually the upper bound of $|\widetilde
R_{1N}^{(2)}(x)|_{\rm (III)}$ in Eq.~(\ref{bound5}) has a $g^N$ dependence
which may give a large remainder for large  $g$. This suggests that we take
\beq
   x\equiv\alpha g^{1/3}N^\gamma,
\label{choice}
\eeq
to compensate such a dependence on $g$. We shall see in
Section~\ref{sec:numerical} this choice indeed gives a faster convergence.

From the above  results it follows that
for  the sequence $\{S_N(x_N)\}$ to converge to the
exact energy eigenvalue it suffices that $x_N=\Omega_N/\omega$ lie
anywhere within the range,
\beq
   C_1N^{1/3+\epsilon_1}\leq x_N\leq C_2N^{1/2-\epsilon_2},
\label{region}
\eeq
where the constants $C_{1,2}$, $\epsilon_{1,2}>0$ are all fixed as $N$ varies,
so that there are actually an infinite set of sequences $\{S_N(x_N)\}$, all of
which converge to the exact answer. In the next subsection we shall refine
the lower bound of the convergence range.

\subsection{Convergence for $\gamma=1/3$}
\label{sec:oneoverthree}

We analyze now the case with scaling index $\gamma=1/3$. Let us set
\beq
   x=\alpha g^{1/3}N^{1/3}.
\label{alpha}
\eeq
To prove the convergence for
\begin{equation}
   \alpha\geq\alpha_c,
\end{equation}
it is sufficient to study
\beq
   \left[\widetilde R_{1N}^{(2)}(x)\right]_{\rm (III)}
   ={g\beta^N\over x^2}\int_{s_c}^\infty{ds\over s}\,G(s)
   \left[{f(w_2)-f(-2)\over w_2+2}\right],
\label{spest}
\eeq
since all other pieces of the remainder in Eq.~(\ref{summary}) are bounded by
\begin{equation}
   \beta^N\simeq\exp\left(-{N^{1/3}\over\alpha^2g^{2/3}}\right)\rightarrow0
\end{equation}
as $N\rightarrow\infty$. $[\widetilde R_{1N}^{(2)}(x)]_{\rm
(III)}$ can be estimated by the saddle point approximation.

Note [as was done in deriving Eq.~(\ref{bound5})] that at $s\ge s_c$ and
at $x$ large, $\lambda=g/(x^3\beta s)$ is always small, so that the
semi-classical estimate Eq.~(\ref{smallg}) can be applied, yielding
\beq
   G(s)\sim{4^{2K+1}\over\sqrt{2}\pi^{3/2}K!}
   \left({x^3\beta s\over g}\right)^{K+3/2}
   \exp\left(-{4x^3\beta s\over 3g}\right).
\label{gs}
\eeq
The saddle point in $s$ is then determined by a compromise between the factor
$1/w_2^{N-1}$ inside $f(w_2)$ which tends to push it towards $s=\infty$ (where
$w_2\sim-1/s\rightarrow0$), and the tunnelling factor
$\exp[-4x^3\beta s/(3g)]$ which tries to prevent it. With the scaling given
in Eq.~(\ref{alpha}) the leading factor in the integrand for large $N$ is
\beq
   \simeq(-1)^N\exp[-N\phi(\alpha,w_2)],
\eeq
and
\begin{equation}
   \phi(\alpha,w_2)
   \equiv
   -{4\alpha^3(1-w_2)^{3/2}\over3w_2}+\log|w_2|,
\label{exponent}
\end{equation}
where we have set $\beta=1$. The saddle point equation is then approximately
given by
\beq
   {\partial\over\partial w_2}\phi(\alpha,w_2)\bigr|_{w_2=w_{2\ast}}=0.
\label{saddle}
\eeq
We studied this equation numerically, observing 1) that $w_{2\ast}$ is a
decreasing function of $\alpha$, and $w_{2\ast}\rightarrow0$ as
$\alpha\rightarrow0$ and $w_{2\ast}\rightarrow-2$ as
$\alpha\rightarrow\infty$; 2) that $\phi(\alpha,w_{2\ast})$ is an increasing
function of $\alpha$ and
\begin{equation}
   \phi(\alpha_c,w_{2\ast})=0,
\label{criticaleq}
\end{equation}
for
\begin{equation}
   \alpha_c=0.5708751028937741\cdots.
\label{alphac}
\end{equation}
(For $\alpha=\alpha_c$, $w_{2\ast}=-0.2429640299735202\cdots$,
$s_\ast=5.703557953364256\cdots$.)

The equations Eqs.~(\ref{saddle}) and (\ref{criticaleq}) are essentially
the same as the ones given in \cite{SZ}.

Property 1) implies that the saddle point is always inside of the
integration region [recall that for $s_c\leq s<\infty$, $-2\leq w_2(s)<0$].
Property 2) shows that $\phi(\alpha,w_{2\ast})>0$ for $\alpha>\alpha_c$,
hence
\begin{equation}
   \left[\widetilde R_{1N}^{(2)}(x)\right]_{\rm (III)}
   \simeq\beta^N\exp[-N\phi(\alpha,w_{2\ast})]\rightarrow0,
   \quad{\rm if}\quad\alpha>\alpha_c.
\end{equation}
At $\alpha=\alpha_c$, convergence still holds due to the factor $\beta^N$ in
front of the integral,
\begin{equation}
   \left[\widetilde R_{1N}^{(2)}(x)\right]_{\rm (III)}
   \simeq\beta^N\simeq\exp\left(-{N^{1/3}\over\alpha_c^2g^{2/3}}\right)
   \rightarrow0.
\end{equation}

Furthermore $\{S_N\}$ converges if
$\alpha=\alpha(N)$ is a sequence such that
$\lim_{N\rightarrow\infty}\alpha(N)=\alpha_\infty>0$ and
\begin{equation}
   \alpha(N)\geq\alpha_c, \quad \forall N.
\end{equation}
This result is of particular interest: in fact, the particular scaling index 
$\gamma=1/3$ is known to follow from the fastest apparent convergence
condition of the delta expansion for the energy \cite{SZ},
\begin{equation}
   x=\alpha_c\,g^{1/3}N^{1/3}+O(N^{-1/3}),
\label{zinn}
\end{equation}
as well as from the principle of minimal sensitivity for the {\em partition
function\/} of anharmonic oscillator \cite{DJ},
\begin{equation}
   x=\alpha_c\,g^{1/3}N^{1/3}+O(N^0).
\label{duncan}
\end{equation}
Using the fact that the coefficients of the sub-leading terms in
Eqs.~(\ref{zinn}) and (\ref{duncan}) are positive \cite{SZ,DJ}, we thus
complete the proof that the delta expansion converges
to the exact energy eigenvalue for either of these choices for $x$. 

Summarizing the result of this subsection, the lower bound for the index for
convergence is now weakened to $x_N\geq x_{Nc}$ where
\begin{equation}
   x_{Nc}\equiv\alpha_c\,g^{1/3}N^{1/3}.
\label{xcn}
\end{equation}
%

\subsection{Uniformity of convergence of the scaled delta expansion and
divergence of $R_N$ for scaling indices outside the convergence domain}
\label{sec:outside}

The uniform convergence in an arbitrary finite domain $g\in[0,G]$
follows if the upper bound for the remainder
[Eq.~(\ref{upperbound}) and Eq.~(\ref{summary})] is an increasing function
of $g$. This indeed holds in general when $C$ is an non-decreasing function
of $g$, such as $C={\rm const.}$ and $C\propto g^{1/3}$.

Similarly, bounds on the remainder are increasing functions of $K$ for
$g$ and $N$ ($>K$) fixed. It follows that convergence is uniform for
$K\leq K_0<N$. This behavior of the bounds suggests also that convergence
is slower for higher energy levels. 

To see that the scaled delta expansion actually diverges for $\gamma>1/2$,
let us consider the first piece of the remainder, $R_{0N}(x)$. We will find
the upper bound in Eq.~(\ref{bound1}) is almost saturated for large $N$.
Consider the difference of the upper bound and $|R_{0N}(x)|$
[see Eq.~(\ref{bound11})].
\begin{eqnarray}
   {c_0x\beta^{N+1}\over\pi}B(1/2,N+1/2)-\left|R_{0N}(x)\right|
   &=&{c_0\beta^{N+1}\over\pi x}\int_0^\infty du\,
   {u^{-1/2}\over(u+1-\beta)(1+u)^{N+1}}
\nonumber
\\
   &<&{c_0\beta^{N+1}\over\pi x}\int_0^\infty du\,
   {1\over u^{1/2}(u+1-\beta)}
\nonumber
\\
   &=&c_0\beta^{N+1}.
\end{eqnarray}
The last quantity is negligible compared to the upper bound for $\gamma>1/2$
and for large $N$. Therefore
\begin{equation}
   \left|R_{0N}(x)\right|\simeq{c_0x\beta^{N+1}\over\sqrt{\pi}N^{1/2}}.
\end{equation}
On the other hand, all other pieces in Eq.~(\ref{summary}) are
bounded by a constant [note when $\gamma>1/2$, $\beta(x)^N\rightarrow1$ as
$N\rightarrow\infty$]. This shows that the total remainder (hence $S_N$ itself)
diverges as
\begin{equation}
   R_N(x)\simeq-{c_0x\over\sqrt{\pi}N^{1/2}},
\label{linear}
\end{equation}
for $\gamma>1/2$ and $N\rightarrow\infty$. It is easy to see that the behavior
Eq.~(\ref{linear}) also holds for $\gamma=1/2$.

The divergence of the delta expansion for $\gamma<1/3$ on the other hand
arises from $|\widetilde R_{1N}^{(2)}(x)|_{\rm (III)}$, since in this
case the factor $\beta^N$ strongly suppresses all other terms of $R_N$
in Eq.~(\ref{summary}). To show how it diverges we apply
the semi-classical approximation Eq.~(\ref{gs}) and make a saddle
point estimation of the $s$ integral as was done in
Section~\ref{sec:oneoverthree}.

The saddle point equation is now approximately given by
\beq
   {\partial\over\partial s}\left[{4x^3\beta s\over 3g}-N\log s\right]=0,
\eeq
where the asymptotic form, $w_2\sim-1/s$, $s\gg 1$, has been used.
This is justified {\it a posteriori\/} from the resulting saddle point:
\beq
   s_\ast\simeq{3gN\over4x^3\beta},
\eeq
which grows indefinitely for $\gamma<1/3$. Substituting this into
Eq.~(\ref{spest}), we find that
\beq
   \left[\widetilde R_{1N}^{(2)}(x)\right]_{\rm (III)}
   \simeq e^{-N}\left(-{3gN\over4x^3}\right)^N.
\label{estimate2}
\eeq
Thus the delta expansion diverges violently with alternative signs
when $N\rightarrow\infty$ with $\gamma<1/3$. We recognize it as a heritage
of the standard large order behavior of perturbative series
(to which the delta expansion is reduced at $x=1$), which is nicely tamed
at a larger scaling index.

Finally, at the critical index Eq.~(\ref{alpha}) but with $\alpha<\alpha_c$,
the discussion of the previous subsection shows that $S_N$ diverges as
\beq
   \simeq(-1)^N\exp[-N\phi(\alpha,w_{2\ast})],
\eeq
where $\phi(\alpha,w_{2\ast})<0$ for $\alpha<\alpha_c$.

\subsection{Comparison with numerical calculation of $S_N$}
\label{sec:numerical}

To corroborate our proof and verify the general features found in
previous sections, we have made a rather detailed numerical study of $S_N$.
The method used is essentially the one described in \cite{DJ} with some
refinement.

We first note that the $N$-th order approximant $S_N$ Eq.~(\ref{delta})
can be written as
\begin{equation}
   S_N(x)=x\sum_{n=0}^N\sum_{k=0}^{N-n}e_{k,n}
   \left({g\over x^3}\right)^n\beta(x)^k,
\label{altern}
\end{equation}
where
\begin{equation}
   e_{k,n}={\Gamma(3n/2+k-1/2)\over\Gamma(3n/2-1/2)\Gamma(k+1)}c_n.
\end{equation}

To compute the delta expansion for anharmonic oscillator [or for the double
well potential, see the discussion below Eq.~(\ref{delta})] to high orders
therefore it suffices to find standard perturbative coefficients for the
anharmonic oscillator $c_n$ to the desired order $N$ and use the relations
above.

We define the perturbative coefficients for the energy eigenvalue and the
moments as:
\beq
   E^{\rm (pert)}=\sum_{n=0}^\infty c_ng^n,\quad
   \langle q^{\ell}\rangle^{\rm (pert)}=\sum_{n=0}^\infty a_n^{(\ell)}g^n.
\eeq
The coefficients $c_n$ and $a_n^{(\ell)}$ then satisfy the following recursion
formula \cite{Cas,Kill}:
\bea
   &&c_{n+1}={4a_n^{(4)}\over n+1},
\nonumber
\\
   &&a_n^{(\ell+2)}={\ell(\ell^2-1)\over4(\ell+2)}a_n^{(\ell-2)}
                 +{2(\ell+1)\over\ell+2}\sum_{r=0}^nc_ra_{n-r}^{(\ell)}
                 -{8(\ell+3)\over\ell+2}a_{n-1}^{(\ell+4)},
\label{recursion}
\eea
with the initial condition
\beq
   c_0= K+{1\over2},\quad a_0^{(0)}=1,\quad a_n^{(0)}=0\quad{\rm for}
   \quad n\ge 1.
\eeq
Eq.~(\ref{recursion}) has been derived by using the Feynman--Hellmann
relation,
\beq
   {\partial E\over\partial g}
   =\left\langle{\partial V\over\partial g}\right\rangle,
\eeq
and the hypervirial theorem
\beq
   {1\over 4}\ell(\ell^2-1)\langle q^{\ell-2}\rangle
   +2(\ell+1)\langle q^\ell(E-V)\rangle
   -\langle q^{\ell+1}V'\rangle=0.
\eeq
Note that this recursion formula is universal (i.e., independent of $g$):
once $c_n$'s hence $e_{k,n}$'s are computed for each $K$, they can be used
repeatedly to study the delta expansion for various $g$ and for different $x$.
Our numerical results for the delta expansion approximants up to
$100$-th order have been produced in this way.\footnote{In \cite{DJ} a
generalization of the recursion formula Eq.~(\ref{recursion}) to the
delta expansion coefficients of $E$ and $\langle q^\ell\rangle$ is used,
which, lacking the universality, is somewhat less convenient than our method
from practical point of view.}
Actually, with our approach, the coefficients $e_{k,n}$ up to $500$-th order
were computed within $1$ hour CPU time on VAXstation 4000/60
(FORTRAN real*$16$ mode).

Our illustrative results here will be limited to the case of the ground state
energy of the anharmonic oscillator. For the numerical calculation of the exact
energy, we used the method proposed in Ref.~\cite{balsa} and we computed
the numerical value with an estimated relative error $\sim10^{-28}$.

\myfigure{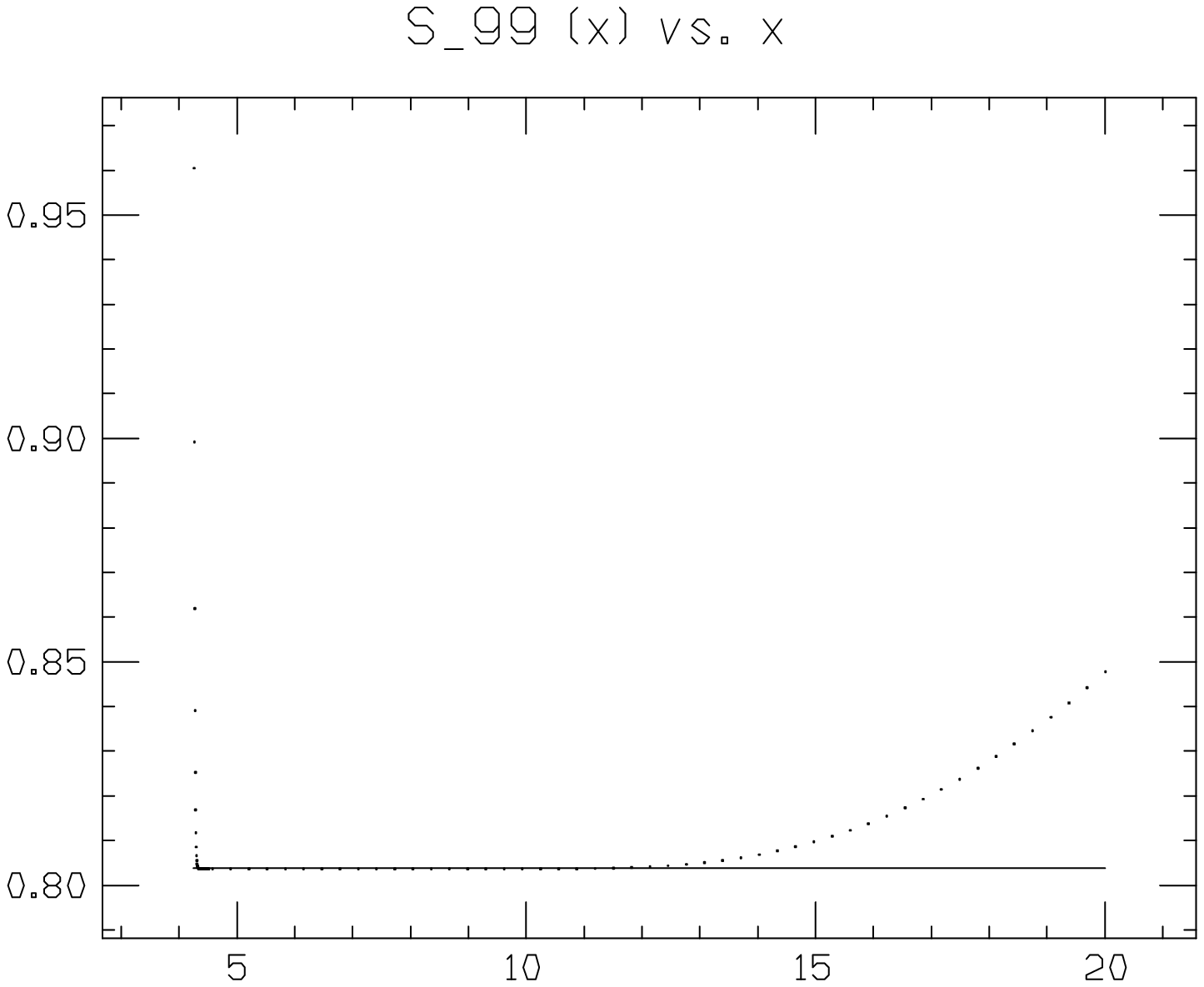}{0.5}
{
Typical plot of $S_N(x)$ ($N=99$, $g=4$). Exact energy (solid line) is
also depicted for comparison.
}
A typical $x$-dependence of the delta expansion approximant $S_N(x)$ at
a fixed order is shown in Fig.~6 for $N=99$, $g=4$. An extremely flat region
in a wide range of $x$, a very sharp rise at the lower end of the plateau,
and a mild increase at large $x$, are clearly seen. These main features
are well explained by the results of Section~\ref{sec:proof} and
Section~\ref{sec:outside}. In particular the position of the sharp left
end edge of the plateau is predicted by our analysis
(Section~\ref{sec:oneoverthree}) to be $\alpha_cg^{1/3}N^{1/3}$ which is
$\simeq 4.19$ for $N=99$, $g=4$.   

\myfigure{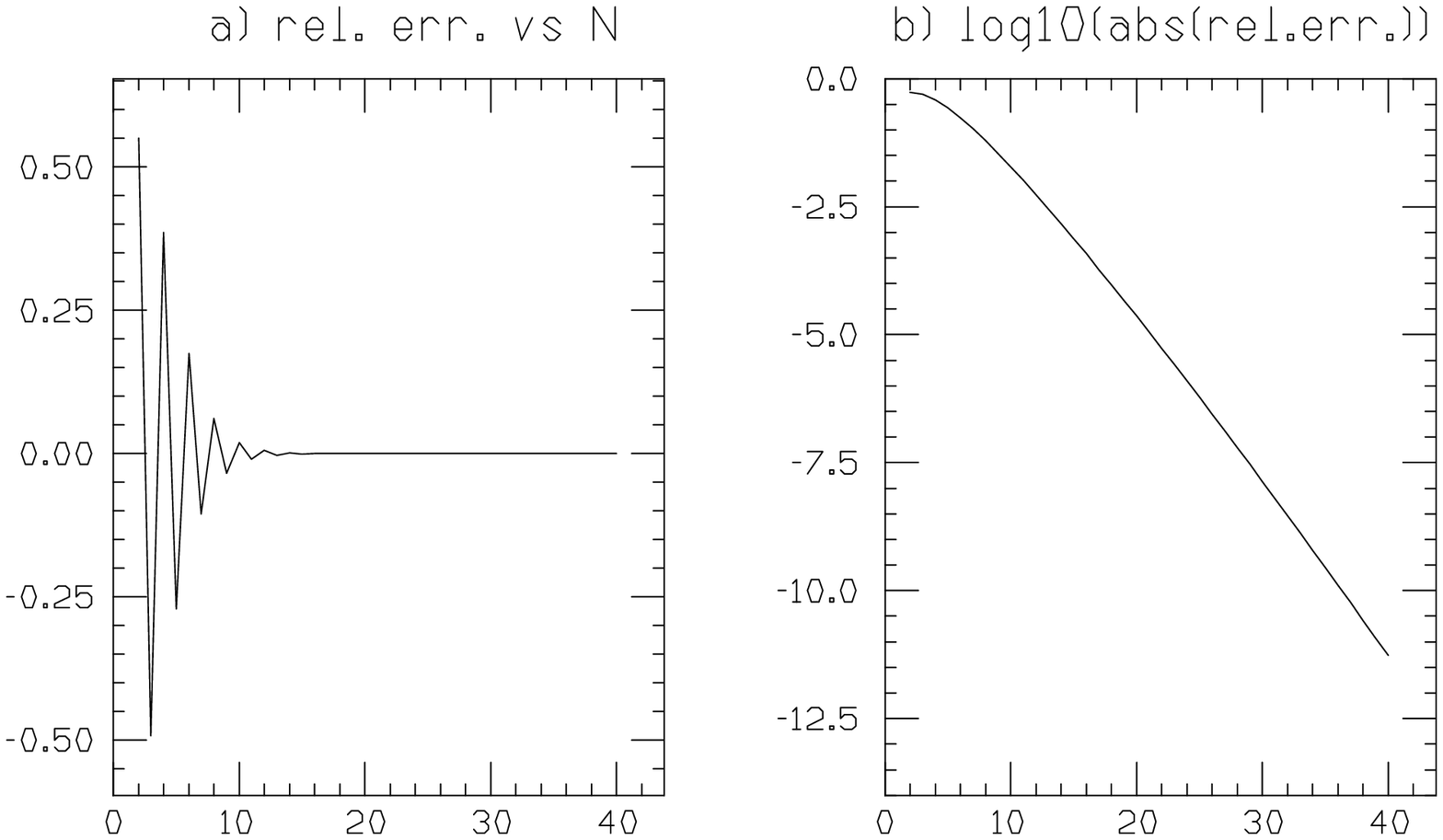}{0.5}
{
a) the relative remainder $\epsilon_N\equiv R_N/E$ with $x=N^{0.35}$
and $g=4$ is plotted versus $N$,
b) same but $\log_{10}|\epsilon_N|$ is plotted.
}
In Fig.~7a and Fig.~7b the relative remainder $\epsilon_N\equiv R_N/E$ with
$x=N^{0.35}$ is plotted versus $N$
(in Fig.~7b $\log_{10}|\epsilon|$ is plotted). They show that the scaled delta
expansion (with an appropriate index) indeed converges rapidly to the exact
answer. Fig.~8 illustrates moreover the general tendency that the convergence
is faster, the smaller the scaling index is, as long as it is within
the convergence domain, as suggested by the $\gamma$ dependence of the upper
bounds in Eq.~(\ref{summary}).
\myfigure{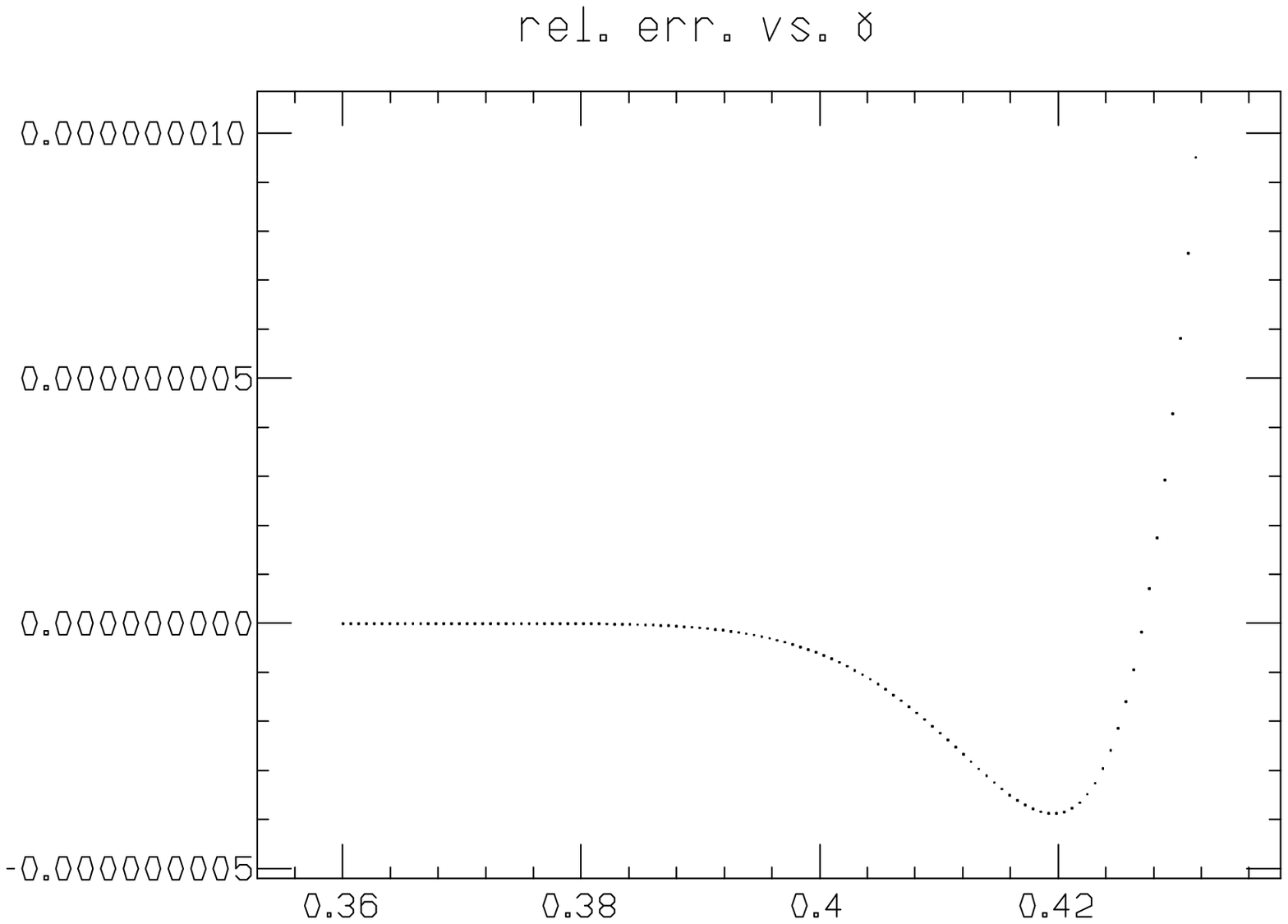}{0.5}
{
Plot of relative reminder $\epsilon_N$
 for fixed $N=40$ and different values of the
scaling index $\gamma$ ($C=1$, $g=4$).
}

Fig.~9 shows a case with the scaling index below $1/3$ ($x=N^{0.3}$). $S_N$ is
indeed seen to oscillate and diverge, as was expected.
\myfigure{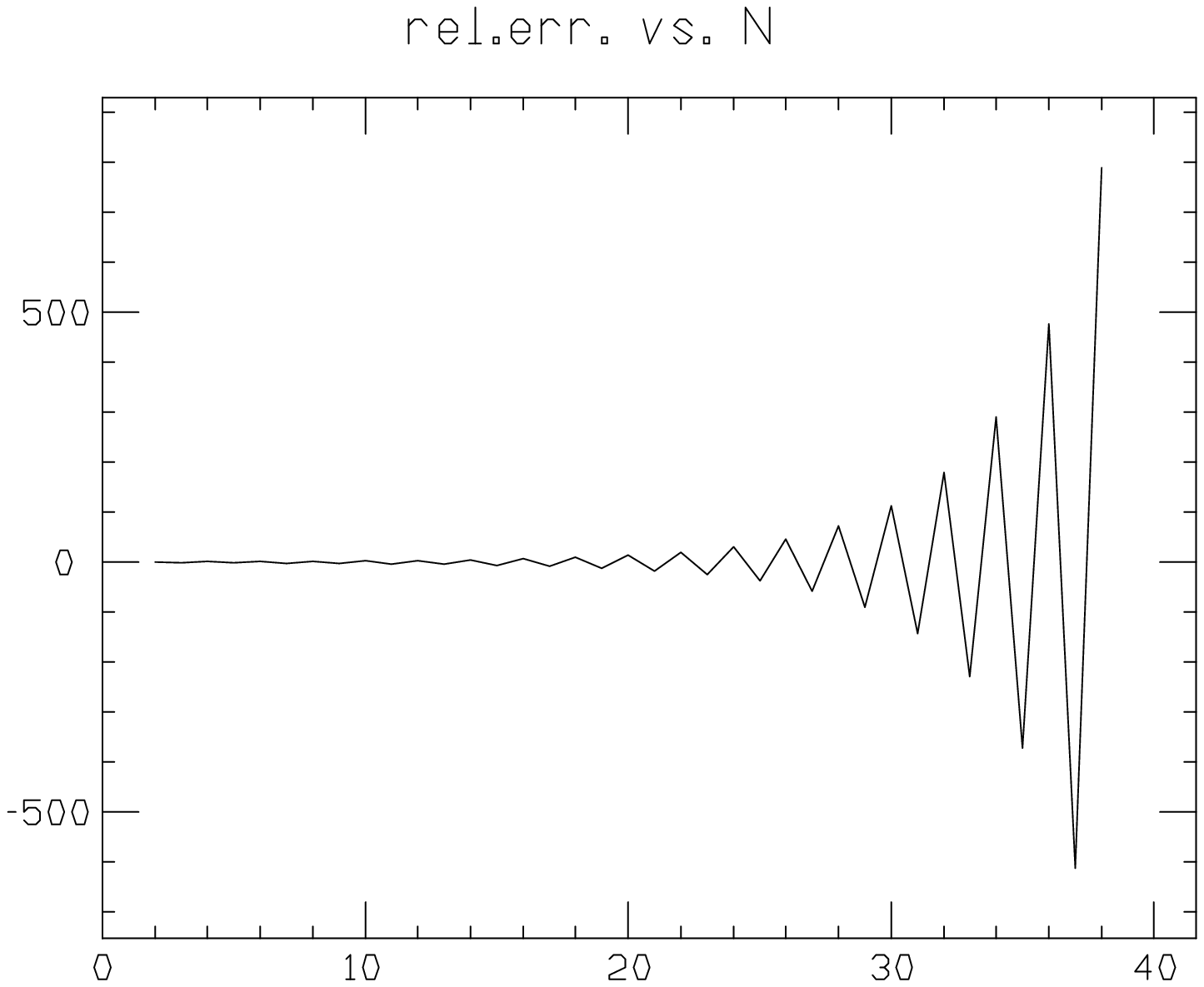}{0.5}
{
Relative reminder $\epsilon_N$ versus $N$ for $x=N^{0.3}$, $g=4$.
}

According to Eq.~(\ref{linear}), for  scaling indices above $1/2$, $S_N$ should
grow as
\beq
   S_N\sim{x\over N^{1/2}}\sim N^{\gamma-1/2}.
\eeq
A logarithmic plot of $\epsilon_N$ (for three different values of
$\gamma$, $\gamma=1$, $2$, $3$) versus $N$ in Fig.~10 shows that such
a simple law is indeed obeyed.

\myfigure{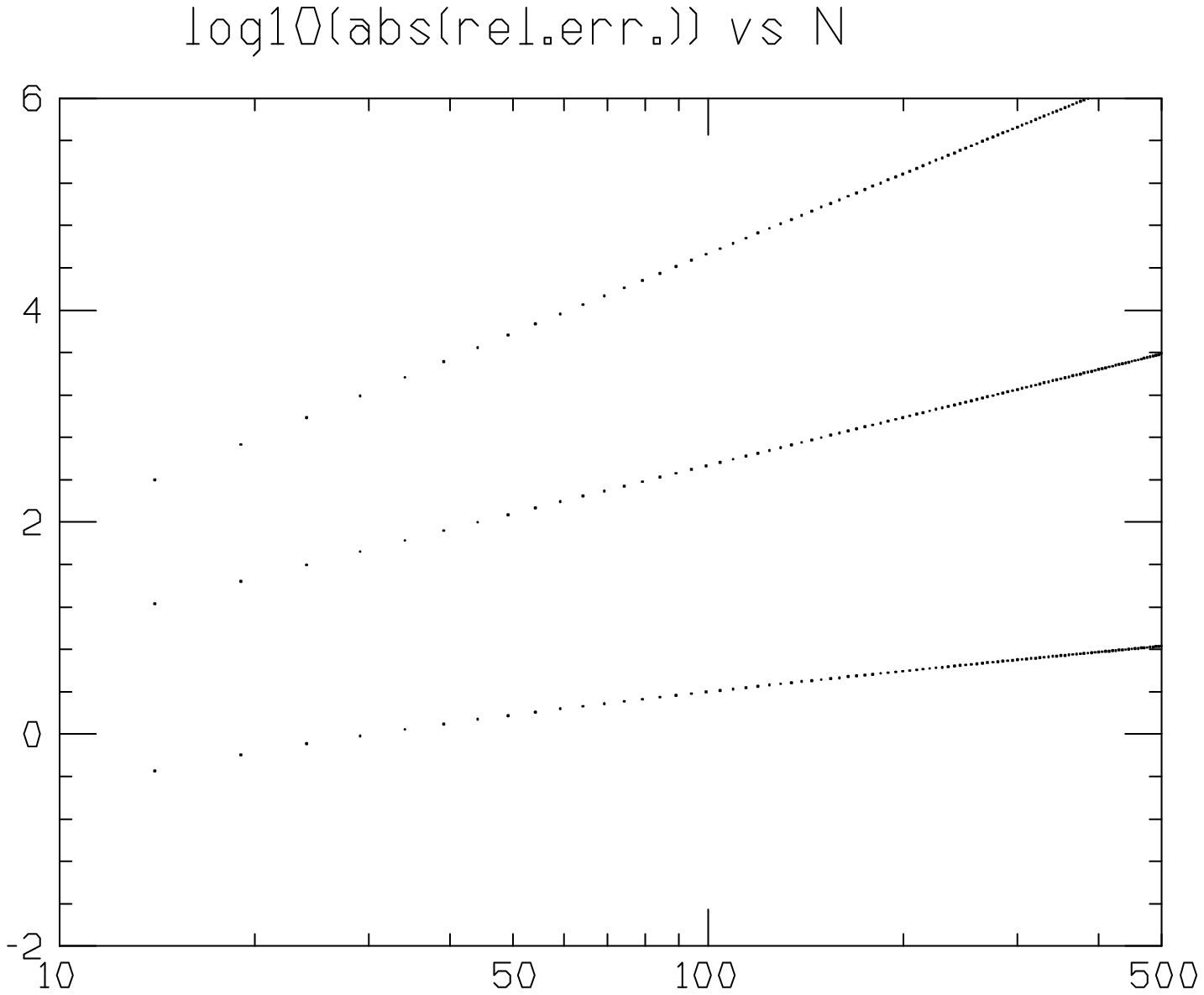}{0.5}
{
Plot of $\log_{10}|\epsilon_N|$ (for three different values of
$\gamma$, $\gamma=1$, $2$, $3$) versus $N$ ($C=1$, $g=4$).
}
The $g$ dependence of $S_N$ at a fixed order $N=40$ is plotted in Fig.~11 for
$\gamma=0.4$. We examined two choices of $C$: $C=1$ (dots) and $C=(g/4)^{1/3}$
(solid line). There is a clear improvement in the second case as we argued
at the end of Section~\ref{sec:proof} [Eq.~(\ref{choice})].
It is also seen that convergence of the scaled delta expansion is slowed at
large $g$, as is suggested from the $g$ dependence of the upper bound in
Eq.~(\ref{summary}).
\myfigure{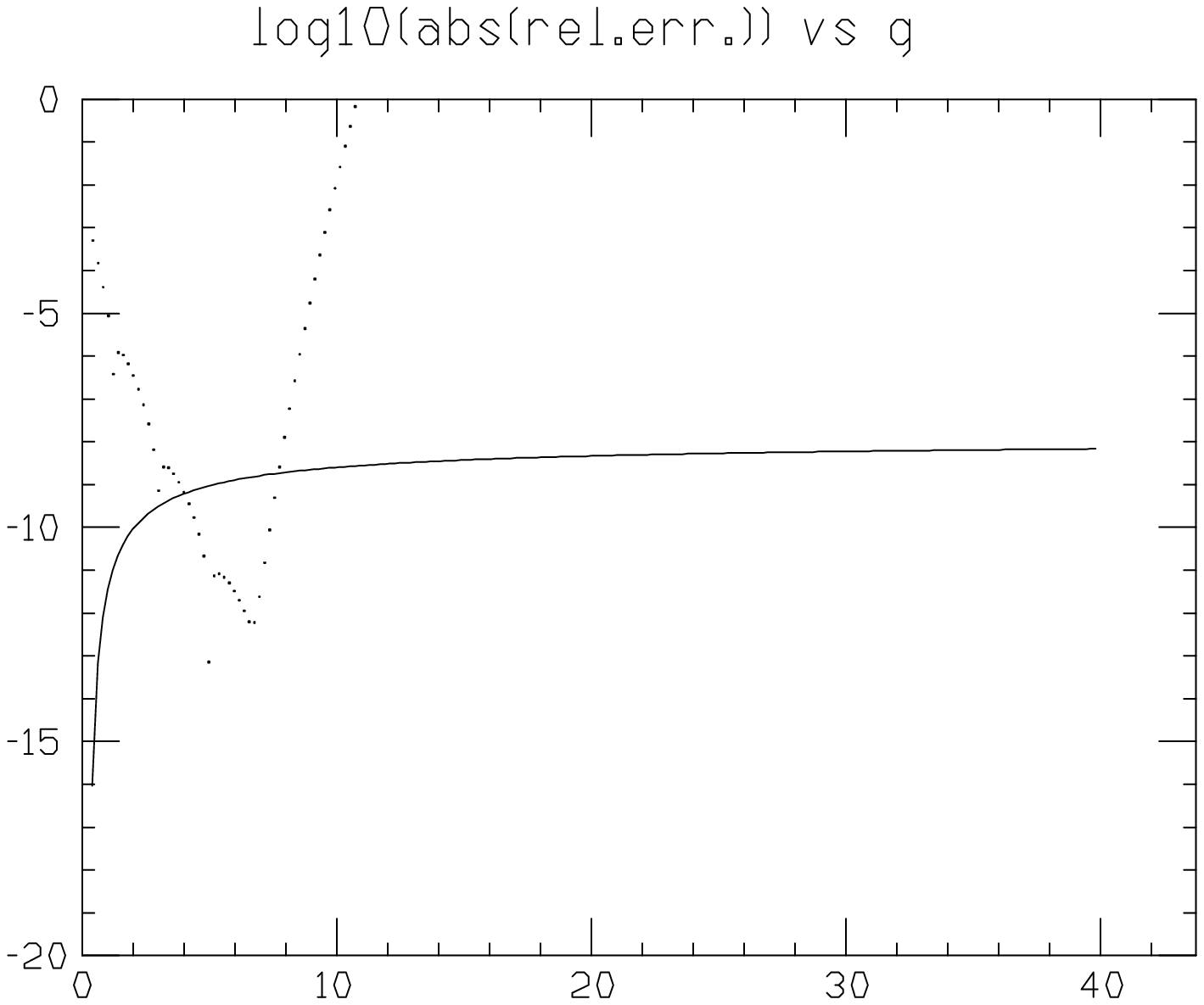}{0.5}
{
Plot 
of $\log_{10}|\epsilon |_N$ versus $g$ at fixed $N=40$, $\gamma=0.4$.
$C=1$ (dots) and $C=(g/4)^{1/3}$ (solid line).
}

All in all, our numerical study of the delta expansion fully confirms the
results of the analyses in Section~\ref{sec:proof},
Section~\ref{sec:oneoverthree} and Section~\ref{sec:outside}.

\subsection{Scaled delta expansion and order dependent mappings}
\label{sec:orderdep}

The order dependent mapping \cite{SZ} was proposed as a resummation
method not based on the Borel transform. This approach involves a change
of the expansion parameter $g$ in Eq.~(\ref{per}) to a new variable $\lambda$
by a conformal transformation:
\beq
   g=\rho F(\lambda),\quad F(\lambda)\sim\lambda+O(\lambda^2).
\label{conformal}
\eeq
The energy is then expanded in $\lambda$ instead of $g$:
\beq
   E=\lim_{N\rightarrow\infty}\widetilde S_N(\rho),\quad
   \widetilde S_N(\rho)\equiv f(\lambda)\sum_{l=0}^N\lambda^lP_l(\rho),
\label{betaexp}
\eeq
with some prefactor function $f(\lambda)$; in the above $P_l(\rho)$ is
a $l$-th order polynomial of $\rho$ with the order $l$. The parameter
$\rho$ in the mapping is then fixed order by order using some criterion
(the exact energy should be $\rho$ independent). In \cite{SZ}, the zero of
$P_N(\rho)$ with the largest module is chosen as $\rho$. This choice
corresponds to the fastest apparent convergence criterion, Eq.~(\ref{fac}).

In the first paper on the order dependent mappings \cite{SZ}, the equivalence
of the linear delta expansion for the anharmonic oscillator (actually they
called $\theta$ instead of $\delta$) and a particular order dependent
mapping was already realized. In~\cite{SZ}, they showed a mapping
\bea
   F(\lambda)={\lambda\over(1-\lambda)^{3/2}},
\label{conf}
\eea
with (in our notation)
\begin{equation}
   \lambda=\beta(x)=1-{1\over x^2},\quad
   \rho={g\over x(x^2-1)}
\end{equation}
is equivalent to the delta expansion. To see this, note that the sum of the
delta expansion up to $N$, $S_N(x)$ in Eq.~(\ref{delta}) can be expressed in an
another form (with $\omega=1$)
\begin{equation}
   S_N(x)={1\over\sqrt{1-\beta}}
   \sum_{l=0}^N\beta^l\sum_{n=0}^lc_n
   {\Gamma(n/2+l-1/2)\over\Gamma(3n/2-1/2)\Gamma(l-n+1)}
   \left[{g\over x(x^2-1)}\right]^n.
\label{anotherform}
\end{equation}
A comparison of $\widetilde S_N(\rho)$ in Eq.~(\ref{betaexp}) with
Eq.~(\ref{anotherform}) shows the above equivalence. The only formal
difference being the choice of the free parameter: $\rho$ in the first case,
and $x$ in the second one.

Our proof of convergence of scaled delta expansion can thus easily be
translated into the one for the order dependent mapping with a scaled parameter
$\rho$. It might be called a ``scaled order dependent mapping.'' For
the scaling $x=CN^\gamma$, one has $\rho\sim g/(C^3N^{3\gamma})$ for 
large $N$. It follows from the proved range of convergence, Eq.~(\ref{index}),
that if $\rho$ is scaled as $\rho=gC'/N^{\gamma'}$  with
\beq
   1<\gamma'<{3\over2},
\eeq
or according to Eq.~(\ref{newindex}), as
\beq
    \gamma'=1,\quad C'\leq{1\over\alpha_c^3},
\eeq
then the order dependent mapping gives a sequence convergent to the exact
answer (vice versa, a divergent one for $\gamma'<1$ or $\gamma'>3/2$). In
particular, the extreme case with $\gamma'=1$; $C'=1/\alpha_c^3$
corresponds to a particular choice made by Seznec and Zinn-Justin \cite{SZ}
(motivated by the fastest apparent convergence criterion).

For the zero dimensional case our convergence proof 
(\ref{sec:appa}) implies that an order dependent mapping based on
\begin{equation}
   F(\lambda)={\lambda\over(1-\lambda)^2},
\end{equation}
converges to the exact answer if $\rho$ is scaled as
$\rho=gC'/N^{\gamma'}$ with
\begin{equation}
   1<\gamma'<2,
\end{equation}
or
\beq
   \gamma'=1\quad{\rm and}\quad C'\leq{1\over\alpha_c^4}. 
\eeq
This follows from a relation $\lambda=\beta(x)$ and $\rho=g/[x^2(x^2-1)]$
which holds in this case (see \cite{SZ}). 

\section{Discussion}
\label{sec:discussion}

Apart from those features specific to  the delta expansion, the convergence
proof presented in Section~\ref{sec:proof} relies essentially only on 
dispersion relation for the energy, Eq.~(\ref{disp}), and on the small coupling
constant behavior of the imaginary part of energy Eq.~(\ref{smallg})
(determined from a semi classical estimate of tunnelling amplitudes).
These are the same ingredients used in the study of the large order behavior
itself \cite{GZ}. We therefore believe that it should be possible to
generalize our proof to a wide class of quantum mechanical as well as
field-theoretic models. The way in which the trial frequency in introduced,
however, might be different case by case.

The lower bound for the scaling index $1/3$ in the case of quantum mechanical
anharmonic oscillator reflects the physical dimension of the coupling constant,
hence the dimensionality of the system $D$ itself. A simple minded
generalization of the proof would yield the lower bound $1/(4-D)$ for the
index, suggesting that the cases of higher dimensional theories ($D\ge2$)
require a nontrivial extension of the method (e.g., modifying the way
$\Omega$ is introduced).

As compared to the conventional approach using the Borel resummation,
the present method is more direct, requiring no analytic continuation of the
Borel transform. However it is still to be proved that all (or some subset of)
cases in which the perturbative series is Borel summable can be treated
by scaled delta expansion.

As regards the interesting cases with degenerate classical minima where
the standard perturbation series is not Borel summable, our preliminary
numerical as well as analytical study on the quantum mechanical double
well has not yet yielded a definitive answer on the applicability of
scaled delta expansion. The convergence proof of Sections~\ref{sec:proof}
and \ref{sec:oneoverthree} clearly does not apply to that case as it stands.
The main difference being
\beq
   \beta_{\rm DW}=1+{1\over x^2}>1
\eeq
now. We hope to come back to this problem in a near future.

Although the scaled delta expansion is quite different in spirit from the
``optimized'' delta expansion in which the trial frequency $\Omega$ is
determined order by order, for instance, by principle of minimum sensitivity,
the results of this paper in no way diminish the virtue of the latter
approach. Rather, we believe that our results put the optimized delta
expansion on a much firmer ground.

The remarkable (empirical) success of principle of minimum sensitivity
for the anharmonic oscillator energy eigenvalues at lowest orders,
might well be related to the proven existence of an infinite set of
sequences that satisfy
\beq
   \alpha_c\,g^{1/3}N^{1/3}\leq x_N\leq C_2N^{1/2-\epsilon_2},
\label{mainresult}
\eeq
all of them converging to the correct answer. This may be of practical
importance because in more complicated systems optimization procedure
might be essential in getting a rapidly converging answer.

Also, as a by-product of our analysis, optimized delta expansion based on
the fastest apparent convergence criterion (with $x_N$ lying just on the
lower bound of Eq.~(\ref{mainresult}) \cite{SZ}), has been given a rigorous
proof of convergence.

Let us compare also the results found here with those of Bender, Duncan and
Jones \cite{BDJ}. In the zero dimensional case, they find that if $x$ is
scaled as $N^{1/4}$ --- corresponding to principle of minimum sensitivity ---
then the delta expansion converges for {\em both\/} signs of $\omega^2$.
(See also~\cite{SZ}.) We proved in \ref{sec:appa} that $S_N$ converges with
$\gamma$ in a much wider range ($1/4<\gamma<1/2$) with an arbitrary
proportional constant or $\gamma=1/4$ and $C\geq\alpha_cg^{1/4}$.
On the other hand it is not obvious in our approach to see the convergence
for $\omega^2<0$ with $\gamma=1/4$. Evidently, the two approaches are
somewhat complementary in this regard.

In the quantum mechanical case, the convergence index $1/3$ (with a
particular proportionally constant) found by Duncan and Jones \cite{DJ}
for the full generating functional at {\em finite temperatures},
is again just on the boundary of the convergence domain
Eq.~(\ref{mainresult}) for energy eigenvalues. Therefore with their
particular scaling [see Eq.~(\ref{duncan})], delta expansion is proved
now to converge to the correct answer for energy eigenvalues as well. 

To conclude, there is still much to be clarified but the scaled delta
expansion appears quite promising as a new resummation method for
the perturbative series. Generalization of our proof to the anharmonic
oscillator Green's functions, as well as to some other simple cases,
will be discussed in a forthcoming  paper.

\bigskip
\noindent
{\bf Acknowledgments:}
We thank T. T. Wu for correspondence and G.~Paffuti and  G.~Viano for
discussions.
H. S. and K. K. wish to thank, respectively, the members of Dipartimento
di Fisica, Universit\`a di Genova and those of Dipartimento di Fisica,
Universit\`a di Pisa, for kind hospitality extended to them.

\appendix
\section{Convergence in the zero dimensional case}
\label{sec:appa}

In this appendix we give a convergence proof for the scaled delta expansion
applied to the integral,
\begin{eqnarray}
   Z(g,\omega)&\equiv&\int_{-\infty}^\infty
   {dq\over\sqrt{2\pi}}\,
   \exp\left[-\left({1\over2}\omega^2q^2+{1\over4}gq^4\right)\right]
\nonumber
\\
   &=&\left({1\over2g}\right)^{1/4}\exp\left({\omega^4\over8g}\right)
      D_{-1/2}\left({\omega^2\over\sqrt{2g}}\right).
\label{exa0}
\end{eqnarray}
The delta expansion is defined by
\begin{equation}
   Z_N\equiv\int_{-\infty}^\infty
   {dq\over\sqrt{2\pi}}\,
   \left\{\exp\left[-{1\over2}\Omega^2q^2
                    -\delta\left({1\over2}(\omega^2-\Omega^2)q^2
                                 +{1\over4}gq^4\right)\right]
   \right\}_N\biggr|_{\delta=1},
\end{equation}
where
\beq
   \left\{f(\delta)\right\}_N
   \equiv\sum_{k=0}^N{f^{(k)}(0)\over k!}\delta^k.
\eeq
As in the one dimensional case, the delta expansion is equivalent to
the substitution
\begin{equation}
   \omega\rightarrow\sqrt{\Omega^2+\delta(\omega^2-\Omega^2)},\quad
   g\rightarrow\delta\cdot g,
\label{eq:integral}
\end{equation}
into the exact expression Eq.~(\ref{exa0}), followed by an expansion
in $\delta$. By using the dispersion relation for $Z$,
\beq
   Z(g,\omega)\equiv{1\over\omega}\widetilde Z(-g/\omega^4),\quad
   Z(g,\omega)={1\over\omega}\int_0^\infty{d\lambda\over\pi}\,
   {\Im\,\widetilde Z(\lambda)\over\lambda+g/\omega^4}
\label{eq:zerodispersion}
\eeq
(where use was made of the large $g$ behavior $Z(g,\omega)\sim g^{-1/4}$),
the same procedure as in the one dimensional case then yields
\begin{equation}
   Z_N=-{\beta^{N+1}\over\pi\omega x}\int_0^\infty{ds\over s}\,
   \Im\,\widetilde Z
   \oint_{C_0}{dw\over2\pi i}{1\over w^{N+1}(w-\beta)}F_0(w,s),
\label{eq:zerostarting}
\end{equation}
where $s\equiv g/(\omega^4x^4\beta\lambda)$ and
\begin{equation}
   F_0(w,s)={(1-w)^{3/2}\over(1-w)^2+sw},\quad
   x\equiv{\Omega\over\omega},\quad\beta\equiv1-{1\over x^2},
\end{equation}
and $C_0$ is a small circle around the origin. This is quite analogous to
$S_{1N}$ considered in Section~\ref{sec:proof}, and the proof is
consequently very similar to that case. We deform the contour so as to
wrap around the pole at $w=\beta$, poles of the function $F_0(w,s)$
and the cut running from $w=1$ to $w=\infty$. The residue of the pole
at $\beta$ gives exactly (the minus of) $Z(g,\omega)$ itself. (We shall
set $\omega=1$ from now on.) The remainder $R_N\equiv Z-Z_N$ is given by
the sum of the contribution of the poles of the function $F_0(w,s)$ and that
of the cut, $R_N=R_N^{\rm (cut)}+R_N^{\rm (poles)}$, where
\begin{eqnarray}
   &&R_N^{\rm (cut)}={\beta^{N+1}\over x}\int_0^\infty{ds\over s}\,
   {\Im\,\widetilde Z\over\pi}
   \int_0^\infty{du\over\pi}\,{1\over(1+u)^{N+1}}{1\over u+1/x^2}
   {u^{3/2}\over u^2+s(1+u)},
\label{eq:zerorncut}
\\
   &&R_N^{\rm (poles)}={\beta^{N+1}\over x}\int_0^\infty{ds\over s}\,
   {\Im\,\widetilde Z\over\pi}
   \sum_i{1\over w_i^N}{1\over w_i-\beta}
   {(1-w_i)^{1/2}\over w_i+1}.
\label{eq:zerornpole}
\end{eqnarray}
From Eq.~(\ref{eq:zerorncut}), we get
\begin{equation}
   \left|R_N^{\rm (cut)}\right|<\beta^{N+1},
\end{equation}
where only the factors $u+1/x^2$ and $u^2$ have been retained in the
denominator of Eq.~(\ref{eq:zerorncut}) and use was made of the relation
$\int_0^\infty ds\,\Im\,\widetilde Z/(\pi s)=c_0=1$, $c_0$
being the zeroth order coefficient of the perturbative expansion.

The poles of the function $F_0(w,s)$ are at:
\begin{equation}
   w_{1,2}\equiv{1\over2}(2-s\mp\sqrt{s^2-4s}).
\end{equation}
When $0\leq s<s_c\equiv4$, the poles $w_1$ and $w_2$ are complex conjugate
to each other and move on the unit circle centered at the origin, $|w_i|=1$.
When $s_c\leq s<\infty$, they are on the negative real axis and
$-\infty<w_1\leq-1$, $-1\leq w_2<0$. To study $R_N^{\rm (poles)}$,
it is convenient to divide the integration over $s$ in three parts.

(I) $0\leq s<s_0$, where $s_0$ ($<s_c$) is arbitrary. In what follows, a choice
$s_0=2$ is made for definiteness. In this range,
\begin{equation}
   \left|\sum_i{1\over w_i^N}{1\over w_i-\beta}{(1-w_i)^{1/2}\over w_i+1}\right|
   <2^{3/4}x^2,
\end{equation}
where use was made of $|w_i|=1$, $|w_i-\beta|\geq1-\beta=1/x^2$,
$|1-w_i|<\sqrt{2}$ and $|w_i+1|>\sqrt{2}$ (note that $w_i=\pm i$ for $s=2$).
The contribution of this part to Eq.~(\ref{eq:zerornpole}) is thus bounded by
\begin{equation}
   \left|R_N^{\rm (poles)}\right|_{\rm (I)}<2^{3/4}x\,\beta^{N+1}.
\label{eq:zerofirst}
\end{equation}

Next to avoid the large residues near $s=s_c$ ($w=-1$) we split the integrand
of Eq.~(\ref{eq:zerornpole}) as:
\bea
   \left|\sum_i{f(w_i)\over w_i+1}\right|
   &\leq&\sum_i{|f(w_i)-f(-1)|\over|w_i+1|}+|f(-1)|
    \left|\sum_i{1\over w_i+1}\right|
\nonumber
\\
   &\leq&\sum_i\max_{\Gamma_i}|f'|
   {L_{\Gamma_i}\over|w_i+1|}+{\sqrt{2}A\over1+\beta}
\label{eq:zeroidentity}
\eea
where $f(w)\equiv(1-w)^{1/2}/[w^N(w-\beta)]$, $L_{\Gamma_i}$ is the arc length
between the points $-1$ and $w_i$, and we have used the fact that
$|\sum_i1/(w_i+1)|\leq A$ ($A$ being a finite constant). Thus the contribution
of the last term in Eq.~(\ref{eq:zeroidentity}) to Eq.~(\ref{eq:zerornpole})
is bounded by
\begin{equation}
   \left|\widetilde R_N^{(0)}\right|
   \leq{\sqrt{2}A\beta^{N+1}\over x(1+\beta)}.
\label{eq:zeronegligible}
\end{equation}
The contribution from the first two terms of Eq.~(\ref{eq:zeroidentity}),
$\widetilde R_N^{\rm (poles)}$, can be bounded as follows.

(II) $s_0\leq s<s_c$. We note $L_{\Gamma_i}/|w_i+1|\leq A'$ with a finite
constant $A'$ and
\begin{eqnarray}
   |f'|&\leq&
   {1\over|w_2|^N}{|1-w_2|^{1/2}\over|w_2-\beta|}
   \left({N\over|w_2|}+{1\over|w_2-\beta|}+{1/2\over|w_2-1|}\right)
\nonumber
\\
   &<&\sqrt{2}\left(N+1+{1\over2\sqrt{2}}\right),
\end{eqnarray}
since $|w_2|=1$, $\sqrt{2}\leq|1-w_2|<2$ and $|w_2-\beta|>1$. Therefore
\begin{equation}
   \left|\widetilde R_N^{\rm (poles)}\right|_{\rm (II)}
   <2\sqrt{2}A'\left(N+1+{1\over2\sqrt{2}}\right){\beta^{N+1}\over x}.
\label{eq:zerosecond}
\end{equation}

(III) $s_c\leq s<\infty$. Note that $L_{\Gamma_i}=|w_i+1|$ in this region.
Also, $|f|$ is a monotonically increasing function for $w_i<0$.
It follows that:
\begin{equation}
   |f'|\leq|f|\left({N\over|w_1|}+{1\over|w_1-\beta|}+{1/2\over|1-w_1|}\right)
   \leq{\sqrt{2}\over1+\beta}\left(N+{1\over1+\beta}+{1\over4}\right),
\end{equation}
where we have used inequalities
$|f|\leq|f(-1)|=\sqrt{2}/(1+\beta)$, $|w_1|\geq1$,
$|w_1-\beta|\geq1+\beta$ and $|1-w_1|\geq2$. Therefore, the contribution of
$w_1$ to Eq.~(\ref{eq:zerornpole}) is bounded as:
\begin{equation}
   \left|R_N^{(1)}\right|_{\rm (III)}
   <\sqrt{2}\left(N+{3\over 4}\right){\beta^{N+1}\over x}.
\label{eq:zerothird}
\end{equation}

On the other hand, we have for $w_2$
\begin{equation}
   |f'|\leq
   {1\over|w_2|^N}{|1-w_2|^{1/2}\over|w_2-\beta|}
   \left({N\over|w_2|}+{1\over|w_2-\beta|}+{1/2\over|w_2-1|}\right)
   <{\sqrt{2}s^N\over\beta}\left(Ns+{1\over\beta}+{1\over2}\right),
\end{equation}
where use was made of $|w_2|>1/s$ (following from $w_2=-(1-w_2)^2/s<-1/s$),
and $1<|1-w_2|\leq2$, $|w-\beta|>\beta$. Since in (III) the argument of
$\Im\,\widetilde Z$ is always small we may use the approximation,
\begin{equation}
   \Im\,\widetilde Z(\lambda)\sim{1\over\sqrt{2}}
   \exp\left(-{1\over4\lambda}\right)\left[1+O(\lambda)\right],\quad
   \lambda\ll1.
\end{equation}
This leads to a bound for the contribution of $w_2$:
\begin{equation}
   \left|R_N^{(2)}\right|_{\rm (III)}
   <{1\over\pi x}\left({4g\over x^4}\right)^N
   \left[{4g\over x^4\beta}N\Gamma(N+1)
         +\left({1\over\beta}+{1\over2}\right)\Gamma(N)\right].
\label{eq:zeroforth}
\end{equation}
From the collection of the upper bounds, Eqs.~(\ref{eq:zerofirst}),
(\ref{eq:zeronegligible}), (\ref{eq:zerosecond}), (\ref{eq:zerothird}) and
(\ref{eq:zeroforth}), we conclude that
\begin{equation}
   R_N\rightarrow0,
\end{equation}
if
\beq
   N\rightarrow\infty,\quad x=CN^{\gamma}\rightarrow\infty,
   \quad{\rm with}\quad1/4<\gamma<1/2.
\eeq

Finally, for $\gamma=1/4$, a saddle point estimate of $[R_N^{(2)}]_{\rm III}$
analogous to Section~\ref{sec:oneoverthree} shows that $R_N\rightarrow0$ if
\begin{equation}
   \alpha\geq\alpha_c=0.972632810758477\cdots,
\label{alphazero}
\end{equation}
where $x\equiv\alpha g^{1/4}N^{1/4}$. On the other hand, the principle of
minimum sensitivity criterion \cite{buck}
or the fastest apparent convergence criterion \cite{SZ,buck} requires 
\begin{equation}
   \alpha=1.072985504616992\cdots>\alpha_c.
\end{equation}
For this choice of $\alpha$ therefore we get an exponential decrease  of the
remainder, in accord with what has been shown earlier \cite{SZ,buck}.

\section{Behavior of the poles of $F(w,s)$}
\label{sec:appb}

In this Appendix the behavior of the roots of Eq.~(\ref{roots})
\beq
   (1-w)^{3/2}+sw=0
\label{s1}
\eeq
which lie on the first Riemann sheet will be studied. A graphical analysis of
\beq
   (1-w)^3=s^2w^2
\label{s2}
\eeq
shows easily that Eq.~(\ref{s2}) has either one (for $0<s<s_c=3\sqrt3/2$) or
three (for $s_c<s<\infty$) real roots. One real root which is always present
and lies in $[0,1]$ (which we call $w_3$ below) however does not satisfy
Eq.~(\ref{s1}). More precisely, $w_3$ is a solution of Eq.~(\ref{s1}) lying
on the second sheet of the cut $w$-plane. Eq.~({\ref{s1}) has then either
two real solutions (for $s_c<s<\infty$) or two complex poles (for
$0<s<s_c=3\sqrt3/2$) in the first sheet, which will be denoted as $w_1$,
$w_2$.

At $s=0$, $w_1=w_2=1$ obviously. Near that point the behavior of $w_{1,2}$ is
\beq
   w_{1,2}=1-s^{2/3}e^{\mp2\pi i/3}.
\eeq

The behavior of $w_{1,2}$ near the confluent point, $-2$, is important in the
analysis of Section~\ref{sec:proof}. By setting
\beq
   s=3\sqrt3/2+\epsilon,\quad w_{1,2}=-2+\delta,
\eeq
in Eq.~({\ref{s2}) and expanding up to the second order of $\delta$ it
is easy to find
\beq
   \delta=\mp{4\over\sqrt[4]{3}}\sqrt{\epsilon}.
\eeq
The two roots thus stay at the symmetric points with respect to $-2$ in
the vicinity of $-2$. The corresponding residues of the function
$F(w,s)$ thus behaves as
\beq
   \Res\,F\biggr|_{w=w_{1,2}}\sim\pm{2\sqrt[4]{3}\over\sqrt{\epsilon}},
   \quad w_{1,2}\sim-2,
\eeq
hence  their sum vanishes when $s\rightarrow s_c$.

The behavior at $s\rightarrow\infty$ can be easily read off from
Eq.~({\ref{s1}): a negative solution of Eq.~({\ref{s1}) should go either
to $-\infty$ ($w_1$) or to $0$ ($w_2$). Their leading behaviors at
$s\rightarrow\infty$ are:
\begin{equation}
   w_1(s)\rightarrow-s^2\left[1+O(1/s)\right],\quad
   w_2(s)\rightarrow-{1\over s}\left[1+O(1/s)\right].
\nonumber
\end{equation}

From Eq.~({\ref{s2}) we see that $w_{1,2}$ satisfy
\beq
   w_1w_2w_3=1,\quad0<w_3\le1,
\label{123}
\eeq
from which we get an interesting property
\beq
   w_1w_2\ge1.
\eeq
When $w_{1,2}$ are complex conjugate to each other ($0\le s<s_c$) this means
\beq
   |w_1|=|w_2|\geq1,
\eeq
which is also obvious from Fig.5.

Actually we can go a little further.  From its implicit definition
Eq.~(\ref{s2})  $w_3$ is seen to satisfy
\beq
   {\partial w_3\over\partial s}=
   -{2sw_3^2\over3(1-w_3)^2+2s^2w_3}\leq0.
\eeq
It then follows from  Eq.~(\ref{123}) that  the product $w_1w_2$
(or $|w_1|^2$ in the complex region) is an increasing function of $s$, 
a fact  also used in Section~\ref{sec:outside}.


\end{document}